\documentclass[pdflatex,sn-vancouver-num]{sn-jnl}

\usepackage{graphicx}%
\usepackage{multirow}%
\usepackage{amsmath,amssymb,amsfonts}%
\usepackage{amsthm}%
\usepackage{mathrsfs}%
\usepackage[title]{appendix}%
\usepackage[table,xcdraw]{xcolor}%
\usepackage{textcomp}%
\usepackage{manyfoot}%
\usepackage{booktabs}%
\usepackage{algorithm}%
\usepackage{algorithmicx}%
\usepackage{algpseudocode}%
\usepackage{listings}%
\usepackage{enumitem}

\usepackage{graphicx}
\usepackage{subcaption}
\usepackage{xspace}

\usepackage[normalem]{ulem}

\setlist[itemize]{itemsep=4pt, parsep=0pt, topsep=6pt}

\usepackage{amssymb}
\usepackage{caption}
\usepackage[nameinlink]{cleveref}
\usepackage{threeparttable}
\usepackage{multirow}
\usepackage{pifont}

\usepackage{makecell}

\theoremstyle{thmstyleone}%
\theoremstyle{thmstyletwo}%

\theoremstyle{thmstylethree}%

\newcommand{\SORT}{\textsc{Sort}\xspace}
\newcommand{\GUPS}{\textsc{GUPS}\xspace}
\newcommand{\KMEANS}{\textsc{KMeans}\xspace}
\newcommand{\SRAD}{\textsc{SRAD}\xspace}
\newcommand{\GEMM}{\textsc{GeMM}\xspace}
\newcommand{\LAVAMD}{\textsc{LavaMD}\xspace}
\newcommand{\RESNETMEDIUM}{\textsc{ResNet50}\xspace}
\newcommand{\BERT}{\textsc{BERT}\xspace}
\newcommand{\RESNETSMALL}{\textsc{ResNet18}\xspace}
\newcommand{\DISTILBERT}{\textsc{DistilBERT}\xspace}

\begin{document}

\newcommand{\Eqref}[1]{(\ref{#1})}

\title{A comprehensive evaluation of spatial co-execution on GPUs 
using MPS and MIG technologies}

\author*[1]{\fnm{Jorge} \sur{Villarrubia}}\email{jorvil01@ucm.es}
\author[1]{\fnm{Luis} \sur{Costero}}\email{lcostero@ucm.es}
\author[1]{\fnm{Francisco D.} \sur{Igual}}\email{figual@ucm.es}
\author[1]{\fnm{Katzalin} \sur{Olcoz}}\email{katzalin@ucm.es}

\makeatletter
\def\titraggedcenter{\leftskip=0pt plus 0.5fil\rightskip=0pt plus 0.5fil\parfillskip=0pt\let\hb=\break}
\makeatother

\affil[1]{\orgdiv{Departamento de Arquitectura de Computadores y Automática} \orgname{Universidad Complutense de Madrid} \orgaddress{\country{(Spain)}}}

\abstract{
To mitigate the increasingly common underutilization of computational resources in modern GPUs, spatial sharing methods enable multiple applications to use them simultaneously. This work presents a comprehensive evaluation of NVIDIA's primary technologies to achieve that goal: Multi-Process Service (MPS) and Multi-Instance GPU (MIG). Our findings reveal a crucial trade-off between MPS's flexibility and MIG's isolation, and reveals many key insights for improving the co-execution strategy according to jobs profiles. In the most favorable scenarios, MPS improves performance by up to 30\% and reduces energy by about 20\%, using its provisioning option to avoid resource monopolization. However, under memory contention, it suffers severe degradation, worsening performance by around 30\%. Conversely, MIG's full hardware isolation resolves memory contention, leading to more consistent improvements, but these gains are tempered by higher overhead, and its rigid scheme can degrade performance in certain cases.

}


\maketitle

\begin{center}
\small\textit{This is a preprint of an article published in 
\textbf{The Journal of Supercomputing}.\\
The final version is available at \url{https://doi.org/10.1007/s11227-026-08312-z}}
\end{center}


\section{Introduction}
\begin{figure}[t]
    \centering
    \includegraphics[width=\linewidth]{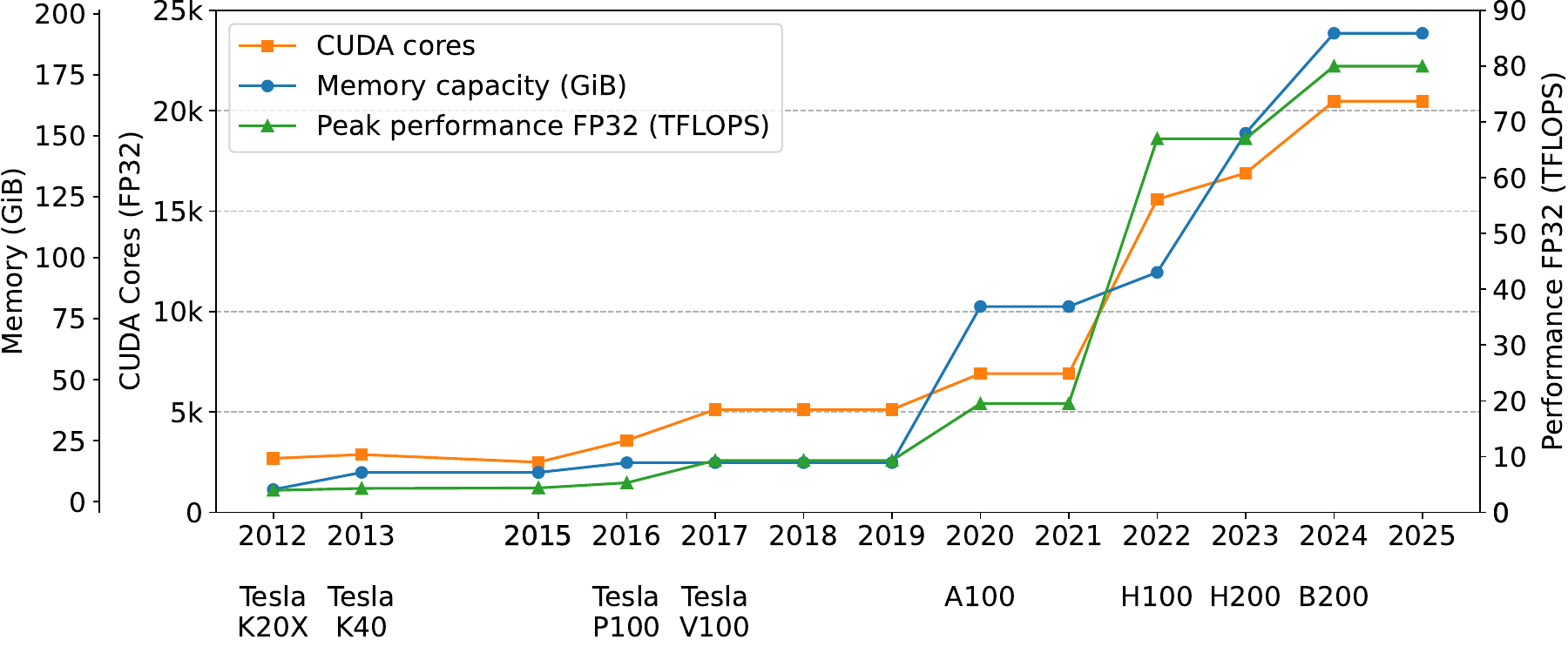}
    \caption{Evolution of FP32 peak performance and amount of resources of the most powerful GPU each year. Data extracted from NVIDIA's official datasheets.}
    \label{fig:evolution_GPU_resources}
\end{figure}

Graphics Processing Units (GPUs) have become the primary devices for accelerating massively parallel computations such as those required for Artificial Intelligence (AI) or a plethora of scientific applications \cite{Pandey2022, Silvano2025}. With the rise of increasingly demanding tasks, notably the training of Large Language Models (LLMs), GPU manufacturers have significantly improved the performance of these devices, particularly by increasing the amount of computational resources available \cite{Atluri2025}. For instance, \Cref{fig:evolution_GPU_resources} depicts the evolution of NVIDIA's flagship GPUs since 2012 and highlights the rise in FP32 cores, DRAM capacity and peak single-precision performance. It is striking that these metrics reveal a particularly pronounced growth from 2019 onwards, coincident with the appearance of the first large-scale LLMs \cite{Radford2019}. For example, until 2019 the NVIDIA V100 was the flagship device, offering 9.3 TFLOPS (in single-precision), 3,584 FP32 cores grouped across 56 SMs ({\em Streaming Multiprocessors}), and 32 GiB of memory with 900 GiB/s bandwidth. By contrast, the current NVIDIA B200 delivers roughly 80 TFLOPS thanks to 20,480 FP32 cores organized into 160 SMs, along with 192 GiB of memory and 8 TiB/s of bandwidth \cite{NVIDIA_Blackwell}.

While this scaling enables effective support for highly demanding applications, many workloads that have traditionally been accelerated on GPUs cannot take full advantage of the vast amount of resources offered by today's datacenter GPUs. Several empirical studies have documented these inefficiencies across a range of High-Performance Computing (HPC) workloads \cite{Adufu2024, Durvasula2024, VillarrubiaFAR2025}, and particularly for Deep Learning~(DL) model inference \cite{Elvinger2025, Gao2024}. This underutilization not only prevents the device from reaching its maximum potential performance but also leads to substantial energy waste \cite{Jie2023}, with significant economic and environmental consequences \cite{Morand2024}. A natural solution to mitigate resource underutilization is the concurrent execution of multiple applications on a single GPU, allowing them to share its computational resources simultaneously. This co-execution approach, known as {\em spatial sharing}, improves upon the default GPU mechanisms for concurrent applications, which solely rely on temporal sharing of the device (i.e., time-slicing, where a scheduler switches between applications). While time-slicing can hide some memory latencies, it fails to address the aforementioned underutilization issues because multiple applications never use the resources simultaneously.

For spatial resource sharing, NVIDIA offers the Multi-Process Service (MPS) \cite{mps} and Multi-Instance GPU (MIG) \cite{mig} technologies, whose main difference lies in their approach towards isolation. Whereas MPS shares resources among applications ---with an option to limit the maximum percentage of SMs, but not to restrict memory system usage--- MIG is based on the physical partitioning of the device into smaller, independent instances, each with its own fully isolated compute and memory resources. Consequently, MPS provides flexible sharing but may suffer from memory-system contention that can violate QoS (Quality of Service) or SLA (Service Level Agreement), while MIG provides a stronger isolation at the cost of coarser execution patterns and more overhead. However, these trade-offs have been insufficiently evaluated in the prior literature \cite{Li2022_MISO, Zhang2024, Zhao2022}, which lacks comprehensive comparisons between both technologies, and especially, in-depth analyses and characterization with diverse workload combinations.

This work contributes to this area by performing a multi-factorial evaluation of different co-execution schemes on modern GPUs. Our main contributions are as follows:
\begin{itemize}
    \item We evaluate the performance of MPS and MIG under different configurations and compare them to baselines such as time-slicing. To this end we run many benchmark combinations (mainly, more computation-oriented or more demanding on memory) that expose the primary strengths and weaknesses of each approach.
    
    \item We analyze the energy consumption of the co-execution schemes considering the power-performance trade-off, and evaluating their scalability with the number of concurrent applications. Specifically, we observe that spatial sharing methods can greatly reduce energy consumption compared to the baselines, and that MPS slightly outperforms MIG in overall efficiency (although in some cases it exhibits significant degradation due to memory contention that MIG does not).
    
    \item We accompany the co-execution experiments with auxiliary metrics on resource utilization, to extract practical lessons. Our main findings include: substantial performance improvements (especially with MIG) when using the appropriate scheme with optimal configuration, severe memory contention issues in MPS, or the benefits of SM provisioning to prevent resource monopolization.
\end{itemize}

The rest of this paper is organized as follows. \Cref{sec:background} provides background on GPU concurrent execution alternatives and supporting technologies. \Cref{sec:methodology} details the experimental setup employed in the evaluation. \Cref{sec:performance} provides a comprehensive performance comparison of the evaluated technologies. \Cref{sec:energy_scalability} evaluates energy efficiency and consumption, as well as scalability with the degree of concurrency. \Cref{sec:related_work} discusses related work. Finally, \Cref{sec:conclusions} summarizes the main conclusions.

\section{Background}\label{sec:background}
This section outlines the fundamentals of time-slicing, MPS and MIG, the primary GPU co-execution technologies evaluated in this work. A brief overview of GPU architecture precedes this description to provide the necessary context.

\subsection{GPU architecture overview}\label{sec:GPU_arch}

\begin{figure}[t]
    \centering
    \includegraphics[width=\linewidth]{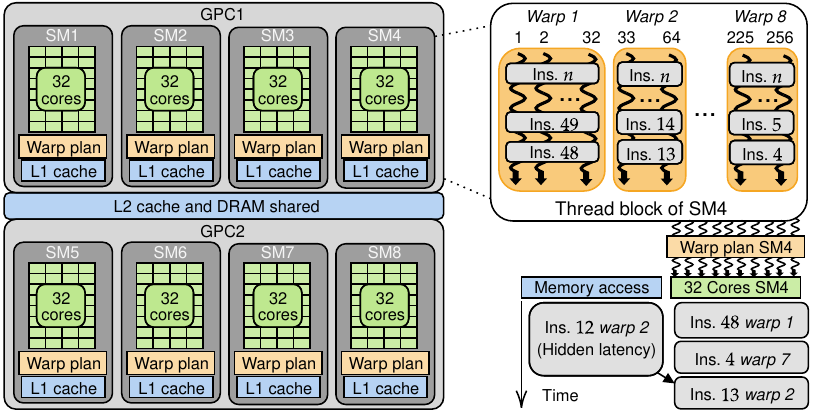}
    \caption{GPU architecture overview (left) and latency hiding by warp switching (right).}
    \label{fig:GPU_architecture}
\end{figure}

A GPU application is composed of distinct pieces of code called kernels, which are launched from the host with a specific configuration of threads and blocks that determines how the work is mapped onto the hardware. Kernels implement a parallel programming model known as Single Instruction, Multiple Threads (SIMT), in which thousands of threads execute the same instructions concurrently but (potentially) on different data \cite{CUDA_guide}.  As illustrated in \Cref{fig:GPU_architecture}, to enable this massive parallelism, GPUs contain thousands of arithmetic logic units, known as cores, which are grouped into physical blocks called Streaming Multiprocessors (SMs), each equipped with its own L1 cache. In turn, several SMs combine to form a Graphics Processing Cluster (GPC), which, in the context of MIG technology, acts as a partitioning unit (see \Cref{sec:MIG_concepts}). Each SM receives an independent thread block and divides it into groups known as warps, performing efficient context switches between these warps to hide memory access latencies. The scheduler not only switches between warps, but several of them are executed simultaneously by mapping their threads to the SM's cores (NVIDIA GPUs use 32 threads per warp, and the number of cores per SM is typically a multiple of this). For each of these warps, all threads execute the same instruction in parallel on different cores (SIMT model). Thus, parallelism is exploited at three levels: different warps execute concurrently on different SMs; within each SM, the instructions for all threads of many warps are executed simultaneously across cores; and within each SM, context is switched between warps to hide memory access latency.


\subsection{Time-slicing}\label{sec:time-slicing}

The most basic method of concurrent execution on a GPU is device time-sharing among multiple applications, which is, in fact, the default behavior when multiple processes are submitted. Technically, the NVIDIA GPU driver time-multiplexes the device among the different GPU processes, known as CUDA contexts, such that each one receives a time slice to execute before the context is switched to another process (see \Cref{fig:no_MPS_scheme}). Historically, these switches were performed at the kernel level, but modern schedulers can now preempt and switch contexts with relatively low overhead (this involves saving the context and incurs a residual cost, such as some pollution of memory structures).

The primary benefit of this time-slicing is fairness among processes, which can particularly reduce the average completion time in multi-application environments. However, it can also yield utilization benefits by eliminating idle periods during operations such as host-to-device (or vice versa) transfers, CPU computations between kernels, or I/O waits. As we will show, our experiments indicate that the aggregate performance of the applications improves slightly with this technique, suggesting that its benefits marginally outweigh the overhead from context switches. While it is possible to modify the scheduling policy and the duration of time slices using tools like Kubernetes with the NVIDIA GPU Operator (e.g., equal-share, fixed-share, long/medium/short slices), we will use the default policy simply as a baseline against which to compare the more advanced methods described below. These methods will be more ambitious than time-slicing, achieving superior improvements in performance and energy consumption, but they require explicit configuration actions that can sometimes be complex (involving scheduling decisions) and incur in a higher overhead.

\subsection{Multi-Process Service (MPS)}\label{sec:MPS}

\begin{figure}[t!]
    \centering
    \begin{subfigure}[b]{0.48\textwidth}
        \centering
        \includegraphics[width=\linewidth]{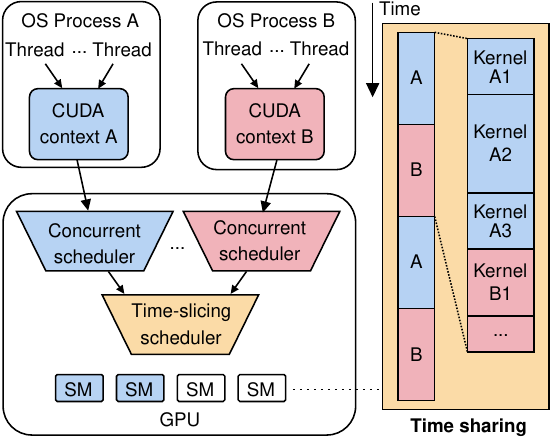}
        \vspace{5pt}
        \caption{Time-slicing co-execution scheme.}
        \label{fig:no_MPS_scheme}
    \end{subfigure}
    \hfill
    \begin{subfigure}[b]{0.48\textwidth}
        \centering
        \includegraphics[width=\linewidth]{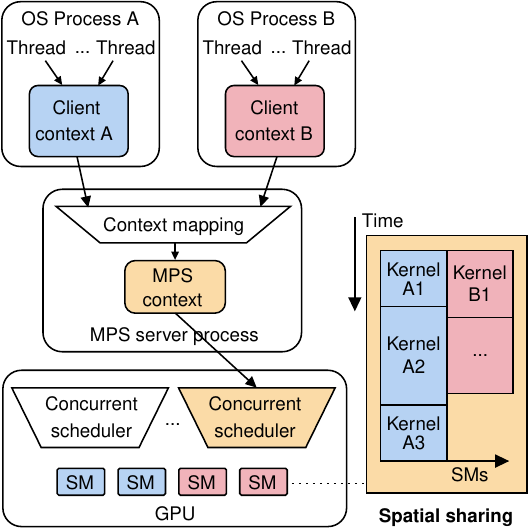}
        \caption{MPS co-execution scheme.}
        \label{fig:MPS_scheme}
    \end{subfigure}
    \caption{Schematic representation of CUDA context management from multiple processes, together with a timing diagram illustrating kernel concurrency.}
\end{figure}

NVIDIA's Multi-Process Service (MPS) allows multiple processes to simultaneously utilize the computational resources of a single GPU. Under normal circumstances, each process that uses a GPU creates its own CUDA context, which encapsulates its entire state (e.g., memory allocations, loaded modules, or address space). However, the GPU driver imposes a fundamental limitation: only one context can be active on the GPU at any given time. As a result, when multiple processes access the same GPU, the hardware defaults to the time-slicing mechanism explained in the previous section, which does not deal with the resources underutilization issues.

MPS overcomes this limitation by implementing a client-server architecture, as depicted in \Cref{fig:MPS_scheme}. An MPS daemon runs as a server, managing a single CUDA context to which multiple client applications connect instead of creating their own contexts. The server funnels the operations from all clients through this unified context, enabling it to dispatch kernels from different processes to the GPU's hardware work queues simultaneously. Underlying technologies, such as Hyper-Q, which equips the GPU with multiple physical queues and a dynamic scheduler, allow these kernels to execute in parallel across the different functional units (spatial resource sharing).

The communication between the MPS client and server is fully encapsulated within the CUDA driver, making MPS transparent to the programmer. The user simply needs to enable the MPS daemon and launch the applications that will share the GPU. However, with this technology, processes compete for shared resources, which can lead to interference. For instance, a process may be delayed because it requires a resource that is being used by another, in addition to other residual conflicts like cache pollution. This can negatively impact performance and may even allow one application to monopolize the device, thereby hindering the ability to provide QoS or SLA.

To mitigate this issue, the Volta generation (released in 2017) introduced several enhancements in MPS, most notably a resource provisioning feature. This allows users to set an upper limit on the percentage of SMs that each process can use. This limit is configured via the \texttt{CUDA\_MPS\_ACTIVE\_THREAD\_PERCENTAGE} environment variable. By using this provisioning, SM sharing can be avoided if the sum of the limits for all concurrent applications does not exceed 100\%. This partitioning ensures that warps from different applications do not run on the same SM, eliminating contention for L1 cache, among other benefits (see \Cref{fig:GPU_architecture}). Nevertheless, the rest of the memory hierarchy (L2 cache, DRAM, and communication buses/crossbars) remains shared and susceptible to conflicts. The official documentation recommends partitioning 100\% of the SMs among the concurrent applications according to their respective demands. In this work, we will evaluate the utility and impact of this provisioning strategy following that advice.



\subsection{Multi-Instance GPU (MIG)}\label{sec:MIG_concepts}

NVIDIA introduced Multi-Instance GPU (MIG) in 2020 to support QoS in the co-execution of GPU applications through isolation. This technology allows a physical GPU to be partitioned into several {\em virtual} sub-GPUs, each with its own dedicated and fully isolated hardware resources. Unlike MPS, which only allows for the provisioning of SMs, each sub-GPU has segregated paths throughout the entire memory system, including independent portions of L2 cache and DRAM, as well as guaranteed bandwidth via dedicated buses and crossbar networks. This physical isolation enables applications to run concurrently on the virtual GPUs without interference. MIG technology is available on datacenter GPUs with the latest architectures: Ampere, Hopper, and Blackwell. Specifically, it is supported by the NVIDIA A30, A100, H100, H200, and B200 models, and recently extended to enterprise and professional GPUs such as RTX 5000/6000 PRO.

MIG division is performed at the GPC level (introduced in \Cref{sec:GPU_arch}), which constitutes the smallest unit that can be isolated. In MIG terminology, these units are called \textit{slices}, a term we will use henceforth. Numerically consecutive slices (where the numbering relates to their physical position on the GPU) can be grouped into what are known as \textit{GPU instances}. These instances are independent and fully isolated virtual devices. The \textit{size} of an instance is determined by the number of slices it comprises, while the set of instances into which the GPU is divided is known as a \textit{partition}.

\begin{figure}[t!]
    \centering
    \begin{minipage}[t]{.6\textwidth}
        \centering
        \includegraphics[width=\linewidth]{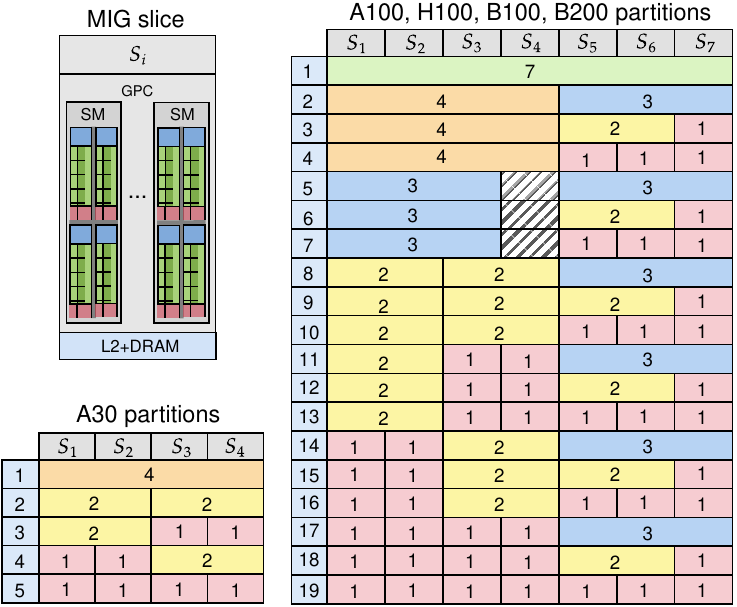}
        \caption{MIG valid partitions.}
        \label{fig:MIG_configs}
    \end{minipage}
    \hfill
    \begin{minipage}[t]{.38\textwidth}
        \centering
      \includegraphics[width=\linewidth]{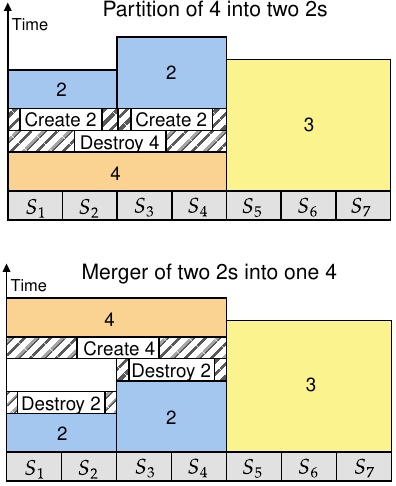}
        \caption{MIG reconfig. examples.}
        \label{fig:reconfig_diagram}
    \end{minipage}
\end{figure}

When partitioning the GPU, not all combinations of slices are valid for creating instances, nor are all instance configurations possible. \Cref{fig:MIG_configs} illustrates the possible configurations for MIG-capable GPUs, which currently follow two partitioning schemes: what we will refer to as the NVIDIA A30 scheme (as it is followed by that GPU, as well as the RTX 5000 PRO and RTX 6000 PRO); and the A100 scheme (used by the NVIDIA A100 and the remaining models). As shown in \Cref{fig:MIG_configs}, GPUs following the A30 scheme consist of 4 slices that can be grouped into instances of size 1, 2, and 4, forming 5 possible configurations. Creating instances of size 3 is not permitted, nor is forming an instance with the two central slices ($\{S_2, S_3\}$). In this scheme, all slices are identical, featuring the same amount of SMs and memory. For example, each slice on an NVIDIA A30 has 8 SMs and 6 GiB of memory (one-fourth of the total GPU resources). In the case of GPUs with the A100 scheme, there are more partitioning possibilities, corresponding to their greater computational power. As illustrated in \Cref{fig:MIG_configs}, these GPUs have 7 slices that can be grouped into instances of size 1, 2, 3, 4, and 7, resulting in 19 possible configurations. As before, instances of size 5 or 6 cannot be formed, nor can slices be arbitrarily grouped to form instances of a valid size. For example, a size-4 instance can only be formed with the first 4 slices: $\{S_1, \ldots, S_4\}$. Additionally, there is a peculiarity where certain configurations with two size-3 instances are possible, which keeps the $S_4$ slice unavailable.

Unlike the GPU partitioning technologies from AMD \cite{mxgpu} and Intel \cite{Intel_SRIOV}, MIG allows for the dynamic reconfiguration of any instance that is idle, i.e., not running a task. Reconfiguration involves destroying certain instances and creating new ones in their place. This process only affects the slices of the modified instances. Therefore, the transformation is transparent and non-disruptive to unmodified instances, which can continue to execute other tasks concurrently. For example, as illustrated in \Cref{fig:reconfig_diagram} (top), the size-4 instance of a 4-3 partition could be destroyed and replaced by two new size-2 instances, transitioning to 2-2-3 without affecting the application running on the size-3 instance. Similarly (bottom of \Cref{fig:reconfig_diagram}), the first instances of a 2-2-3 partition could be merged to transition to 4-3.

\section{Experimental setup}\label{sec:methodology}

\subsection{Experimental environment}

To evaluate the presented co-execution technologies, we primarily used an NVIDIA A30 GPU with 24 GiB of memory, although we also had access to a 40 GiB NVIDIA A100 PCIe to verify the results of some experiments. The host system is equipped with an Intel Xeon Silver 4314 CPU @2.40 GHz and Debian GNU/Linux 12, using NVIDIA driver version 580.82.07 and CUDA 13.0. To ensure fair and reproducible results, all experiments were conducted with persistence mode enabled, locking the memory and SM clock frequencies to 1215 MHz and 1440 MHz, respectively (the maximum supported by the A30). We also verified that no throttling events (clock reduction due to temperature or power limits) occurred by monitoring reports from the Data Center GPU Manager (DCGM) tool~\cite{NVIDIA_DCGM}.

\subsection{Co-execution schemes}\label{sec:schemes}

The main evaluation consists of a comparison between four co-execution schemes based on the three technologies described in \Cref{sec:background}: time-slicing, MPS, and MIG. Since MPS offers a provisioning option, we evaluate this technology both with and without this feature, particularly in the initial experiments. The four schemes evaluated in this study are as follows:
\begin{itemize}
    \item \uline{Time-slicing}: The default temporal switching performed by the scheduler when it receives multiple applications simultaneously (see \Cref{sec:time-slicing}).
    
    \item \uline{MPS}: Spatial sharing with the basic version of MPS, not using the provisioning option (see \Cref{sec:MPS}).
    
    \item \uline{MPS-best-aprov}: Spatial sharing with MPS, but provisioning the maximum percentage of resources that each application can use. This provides a degree of isolation when the total allocation is under 100\% (see \Cref{sec:MPS}). We select the optimal provisioning (\textit{-best-aprov}) by sweeping through resource distributions in 10\% increments, choosing the split that minimizes the makespan (i.e., the completion time of the last application to finish).
    
    \item \uline{MIG}: Spatial sharing with full resource isolation via MIG hardware partitioning (see \Cref{sec:MIG_concepts}). The instances sizes used are detailed in each experiment; however, on the A30, the options are limited, and we will primarily use the 2-2 in \Cref{sec:performance}, and also 1-1-1-1 in \Cref{sec:energy_scalability}.
    
\end{itemize}

\subsection{Evaluation metrics}\label{sec:metrics}

This study focuses on both performance and energy consumption. Performance is evaluated using metrics derived from application execution times under each scheme. Specifically, we use the following:
\begin{itemize}
    \item \uline{Individual slowdown}: The ratio of an application's co-execution time compared to its standalone execution time. In \Cref{sec:performance}, we focus on the individual slowdown of each application to ensure that one application's performance does not improve significantly at the expense of another, which would indicate an unfair or imbalanced resource distribution.

    \item \uline{Overall slowdown}: In \Cref{sec:energy_scalability}, we use the overall slowdown, defined as the total co-execution runtime (i.e., makespan) relative to the sum of the all standalone execution times.
\end{itemize}

This work will analyze energy consumption by examining the fundamental trade-off between performance and power draw, dissecting the independent contribution of each factor. While higher performance reduces execution time, it often requires a power increase. In this context, we measured the total Joules consumed during the execution with the \texttt{nvmlDeviceGetTotalEnergyConsumption} function from the NVML library~\cite{nvidia_nvml}. This value is also divided by the total execution time to calculate the power (in watts), which will be important in the analysis, and is also presented in relation to the values measured during the standalone execution of the applications.

Finally, to understand and analyze the results, these primary metrics are supplemented by auxiliary metrics on resource utilization. One option is the throughput metrics offered by Nsight Compute (NCU), which include peak bandwidth utilization percentages for different memory levels and the achieved percentage of peak SM performance for each kernel (it will be used in \Cref{fig:app_contention_prof} and \Cref{tab:partition_problems}). However, NCU is not compatible with the MPS service because the unified MPS context creates ambiguity in how the tool has to take and organize hardware counters, so we cannot use it in most of our experiments (though we will use it in a few cases where it provides particularly illustrative insights). As other studies on MPS \cite{Elvinger2025, Robroek2024, Weaver2024, Zhao2023MuxFlow}, we will instead use the memory and compute (SM) utilization percentages provided by the DCGM tool for this purpose~\cite{NVIDIA_DCGM}. This tool performs periodic measurements (every 200 ms in our case) to calculate resource utilization. Each period is divided into 100 slots, and by checking for resource activity in each slot, a utilization percentage is obtained. Finally, these percentages are averaged over time.

\subsection{Workloads}\label{sec:meth:workloads}

\begin{table*}[t!]
\centering
\caption{Specification of the applications used for evaluation. The first six belong to the Altis suite \cite{Altis}.}
\label{tab:workloads_specification}

\footnotesize
\renewcommand{\arraystretch}{1.4}
\setlength{\tabcolsep}{3pt}
\begin{tabular}{cm{7.4cm}cc}\toprule
 Test name & \multicolumn{1}{c}{Description} & Problem size & Passes \\ \midrule
 \SORT & Update for modern GPUs of the test from SHOC suite \cite{SHOC} that uses the radix sort algorithm on integer key-value pairs. & Level $2$ & 3\\
 \GUPS & Giga UPdates per Second  measures the GPU's performance in updating randomly generated memory locations. & Level $3$ & 58\\
 \KMEANS & Test 11 different GPU implementations of the popular clustering algorithm. & \makecell{$131072$ points\\$32$ dimensions\\$32$ centroids} & 2\\
 \SRAD & Solves differential equations for image noise reduction. It is a Rodinia \cite{Rodinia} test improved by CUDA Cooperative Groups. & Level $4$ & 1\\
 \GEMM & Square matrix multiplication with the GeMM cuBLAS routine, using medium, single, and double precision. & Level $3$ & 4\\
 \LAVAMD & Calculates the interaction forces between a set of particles and their resulting relocation in 3D space. & Level 3 & 2\\[0.8em]
 \RESNETMEDIUM \cite{ResNet} & A 50-layer Convolutional Neural Network (CNN) for classifying images of size $224\times224$. It has $\sim 25$M parameters.& \makecell{CIFAR-10 (test 10K)\\Batch size 32} & 1 \\[1em]
 \BERT \cite{BERT} & Transformer model used for embedding extraction and text classification. It has $\sim 110$M parameters.& \makecell{Data: GLUE SST-2\\Batch size 32\\Max. sequence 512} & 13 \\
\bottomrule
\end{tabular}
\end{table*}

Regarding the workloads, we selected various tests from the Altis GPU benchmark suite \cite{Altis}, which has been widely used in other evaluations of the literature \cite{DeLaCalle2025, Sainz2022, VillarrubiaDRL2026}. Most of them are configured by a \textit{level} parameter from 1 to 5, that adjusts the input configuration to vary the computational demand ($level\!=\!1$ uses the least demanding configuration and $level\!=\!5$ the most). In other cases, that configuration can be determined manually. \Cref{tab:workloads_specification} lists the chosen applications (with a brief description of each), the level or problem size used, and the number of passes set for each test. Correspondingly, \Cref{tab:app_metrics} shows their standalone execution times in an NVIDIA A30, along with SM and memory utilization metrics (average and maximum), and their energy and power consumption when run alone with full GPU access. It should be noted that for the benchmarks and workloads used in this study, the program that launches the test automatically selects the thread configuration (number of blocks and threads per block for deployment on the GPU as explained in \Cref{sec:GPU_arch}) based on the problem size and the GPU's characteristics.

\begin{table*}[t!]
\centering
\caption{Duration, compute and memory utilization, and energy consumption metrics on an NVIDIA A30 for the applications evaluated for co-execution.}
\label{tab:app_metrics}

\footnotesize
\renewcommand{\arraystretch}{1.2}
\setlength{\tabcolsep}{4pt}
\begin{tabular}{cccccccccc}\toprule
& & & \multicolumn{2}{c}{SMs utilization (\%)}& & \multicolumn{2}{c}{Mem. utilization (\%)}
  & & \\
\cmidrule{4-5}\cmidrule{7-8}
 Type & Test name & Time (s) & Avg. & Max. & & Avg. & Max. & Energy (J) & Power (W) \\ \midrule
  M & \SORT & 5.16 & 36.62 & 70 & & 24.60 & 52 & 321.62 & 62.33 \\
  M & \GUPS & 4.81 & 17.31 & 54 & & 17.72 & 67 & 337.18 & 70.10 \\
  M & \KMEANS & 5.31 & 36.92 & 75 & & 23.92 & 45 & 304.52 & 57.35 \\
  M & \SRAD & 6.04 & 38.14 & 69 & & 24.08 & 40 & 367.29 & 60.81 \\ \midrule
  C & \GEMM & 5.74 & 78.42 & 100 & & 7.87 & 35 & 543.18 & 94.63 \\ 
  C & \LAVAMD & 6.21 & 34.41 & 100 & & 0.61 & 6 & 531.51 & 85.59 \\ \midrule
  RDL & \RESNETMEDIUM & 5.36 & 65.42 & 85 & & 25.10 & 42 & 540.11 & 100.77 \\
  RDL & \BERT & 5.30 & 45.74 & 78 & & 10.76 & 25 & 526.95 & 99.42 \\
\bottomrule
\end{tabular}
\end{table*}

As can be observed, the average SM utilization is quite low (below 40\%) in most cases, which indicates sparse resource usage, making them good candidates for co-execution. The exception among the Altis applications is \GEMM, with 78\% average SM utilization. While this might suggest it is not a good candidate for co-execution, its memory utilization is very low (7.87\% on average), leaving memory resources free for other applications. We wanted to include such an application because it represents a different profile and, as we will see, its results will be quite interesting.

To ensure the results of some experiments were illustrative, we sought to balance the application runtimes without altering their inherent nature. This is because if one application finishes much later than another, it spends a significant amount of time running alone, which dilutes the co-execution results. This effect is particularly pronounced with approaches like MPS, where the remaining application can expand to use all GPU resources once the other has completed. While one can imagine applications in a continuous co-execution stream, for practical purposes, we must evaluate them over a finite horizon. Thus, the number of execution passes was adjusted so that the test durations were similar; \Cref{tab:workloads_specification} specifies the passes for each, while \Cref{tab:app_metrics} shows that the execution times are all within a narrow range of 5 to 6 seconds.

For our analysis, it is important to discriminate how each application uses memory, as this is the resource where MIG and MPS primarily differ (the first isolates it at all levels, while the latter does not). Accordingly, the selection was guided by the goal of having applications that were sufficiently differentiated in this respect, which we classify into two types (\Cref{tab:app_metrics} indicates the type for each):
\begin{itemize}
    \item \uline{Type M} (Memory): These applications show moderate memory utilization in \Cref{tab:app_metrics} (average values around 20\%), which aligns with other profiling metrics from the original Altis suite paper \cite{Altis}.

    \item \uline{Type C} (Compute): The memory utilization of these applications is residual (values of just 8\% and almost nil for LavaMD), meaning their resource demand is predominantly computational. These results also align with profiling metrics from the suite's presentation \cite{Altis}.
\end{itemize}

Although the Altis tests correspond to real applications, they are designed to test the GPU in a particular way, and limiting our study to them could restrict its validity. In this context, we have also included two widely used Deep Learning applications: inference with ResNet50 and BERT models, using the TorchVision library within the PyTorch framework. Their evaluation datasets and batch sizes are specified in \Cref{tab:workloads_specification}. We classify these models into a new category, \uline{Type RDL} (Real Deep Learning), which we will use primarily for additional validation. As shown in \Cref{tab:app_metrics}, BERT makes moderate use of SMs and low use of memory, whereas ResNet50 is quite intensive in both. These distinct behaviors will serve to corroborate the observations obtained from the Altis benchmark tests.

Finally, it is worth noting that the power consumption values in \Cref{tab:app_metrics} are well below the 165W Thermal Design Power (TDP) of the NVIDIA A30. The applications that come closest are the RDL workloads and GeMM, likely due to their higher intensity, yet they barely reach 100W, while other cases consume only around 60W. These data will serve as baselines for analyzing relative changes in consumption, which, as indicated, never approached the TDP causing throttling effects.

\section{Performance analysis}\label{sec:performance}


This section presents the results of pairwise co-execution experiments. While the next section examines concurrency beyond two applications, restricting the scope here to pairs enables an exhaustive exploration of all combinations and makes it easier to attribute performance interactions to specific patterns. For each pair, we compute the individual slowdown for both applications to detect and prevent imbalances: a performance gain for one job can hide a severe loss for its partner, and such asymmetries are not well captured by overall slowdown or by the slowdown of the first finishing job. As previously noted, runtimes were adjusted so standalone durations are comparable to reduce bias from imbalanced executions (see \Cref{tab:app_metrics}).

\subsection{Impact of memory resource contention}

\begin{figure}[t]
    \centering
    \includegraphics[width=\linewidth]{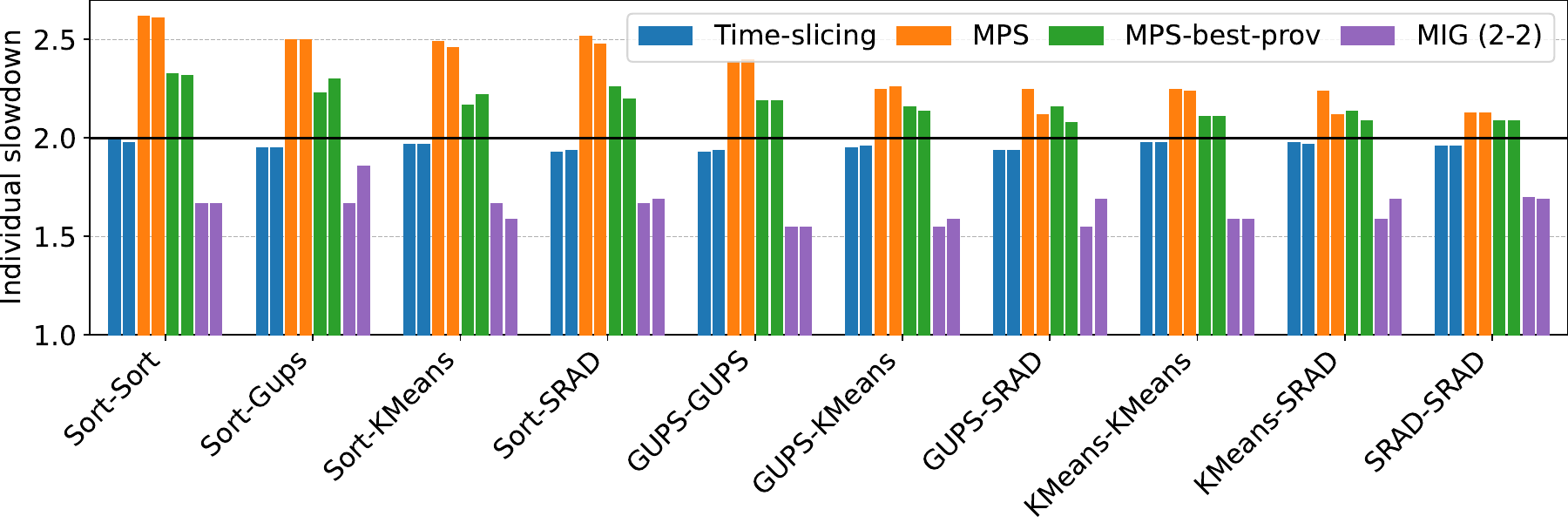}
    \caption{Co-execution slowdown for M-type applications pairs (one bar per application).}
    \label{fig:slowdown_MM}
\end{figure}

\begin{figure}[t]
    \centering
    \includegraphics[width=\linewidth]{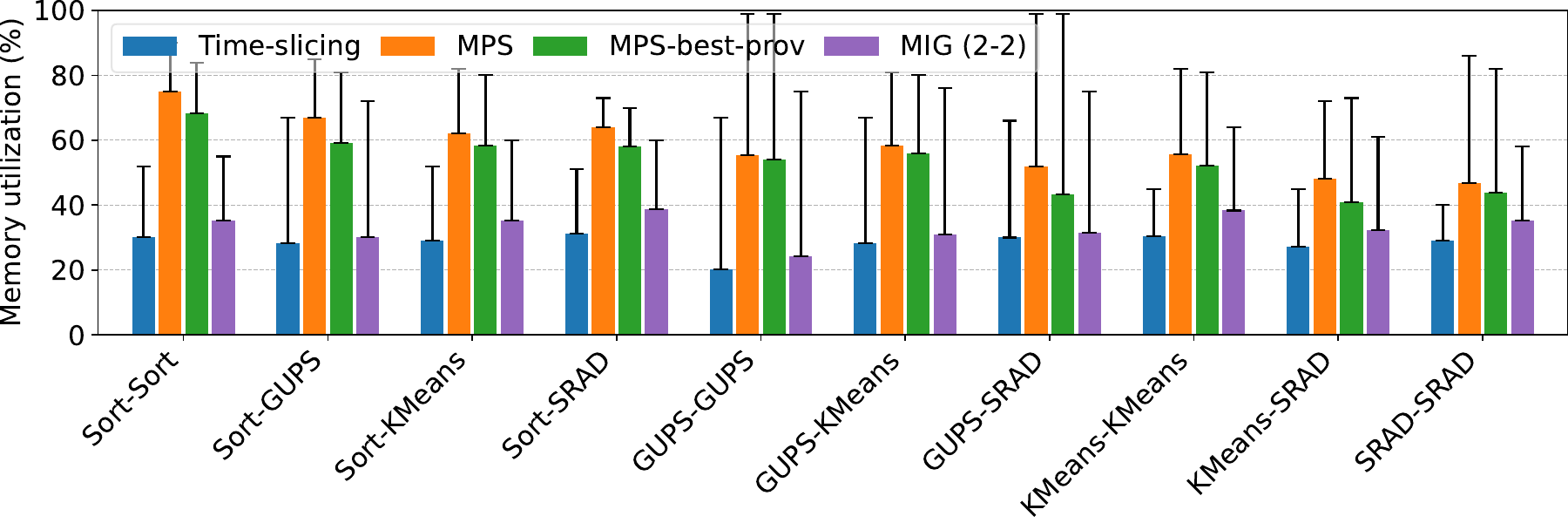}
    \caption{Average memory utilization \% (and vertical projection to maximum) for M-type application pairs.}
    \label{fig:memory_utilization_MM}
\end{figure}

\Cref{fig:slowdown_MM} compares the individual slowdown of the co-execution schemes for each pair of M-type applications. For reference, the black horizontal line marks a slowdown of $2\times$, which can be considered a satisfactory reference, since it means that the application runs concurrently with another without either exceeding twice its solo execution time. It is particularly striking that MPS-based schemes perform so poorly for these combinations: they systematically degrade both applications' performance, reaching slowdowns of up to 2.5 (especially for pairs including \SORT). The underlying cause is revealed in \Cref{fig:memory_utilization_MM}, which illustrates a considerable surge in memory utilization with MPS compared to time-slicing or MIG, evident in both average and peak values. In cases of greater slowdown (\SORT test), it can be observed that MPS exhibits an average memory utilization of roughly 60\%, rising to nearly 80\% for the \SORT–\SORT pair (the highest slowdown), with sustained 100\% peaks. By contrast, time-slicing barely exceeds 20\%—values that closely match the solo-execution metrics in \Cref{tab:app_metrics}—while executions under MIG show only a slight increase over those solo values. This reflects prolonged DRAM engagement, which points to contention in the GPU memory subsystem shared under MPS. The likely causes are an increased rate of misses in lower-level caches and simultaneous access attempts saturating the L2 cache and DRAM bandwidth\footnote{As previously mentioned, finer-grained profiling, such as that offered by NCU tool, is unfortunately incompatible with MPS, preventing analysis of lower-level metrics like cache misses.}. In the rest of the executions, the differences in memory usage are not as exaggerated, but still very noticeable, which justifies the performance degradation observed in the slowdown.

While some of these effects were foreseeable given MPS's shared-resource design, their impact on performance is severe and it is concluded that MPS should be avoided for these kinds of workloads. Note that even the default time-sharing mechanism, which only switches between kernels and can just marginally hide latencies at the cost of a small preemption overhead, achieves slowdowns slightly below 2$\times$ in \Cref{fig:slowdown_MM}. In contrast, those of MPS are well above this, as we have highlighted. Furthermore, the performance difference between MPS and MIG reaches almost a full slowdown point in many cases (i.e., approximately 100\% of the solo execution time). MIG, with a partition of 2 slices for each application (2-2), yields quite good performances with slowdowns below 2$\times$ in all cases and approaching 1.5$\times$ in some, indicating that these applications are highly suitable for spatial co-execution, as they are far from saturating the GPU, but require the strong resource isolation that MIG provides and MPS lacks.

Although the co-execution performance under MPS-best-prov in \Cref{fig:slowdown_MM} remains poor, it represents a slight improvement over basic MPS. Recall that if the sum of the SM percentages does not exceed 100\% (as in this case in MPS-best-prov as explained in \Cref{sec:schemes}), isolation is achieved at the SM level. This means no single SM hosts threads from both applications, which prevents contention for shared resources like the L1 cache. While this could potentially be counterproductive for warp interleaving, the thread and block configurations for each application are generally sufficient to achieve 100\% theoretical occupancy (as verified with NCU tool for numerous kernels). \Cref{fig:memory_utilization_MM} shows a slight decrease in memory utilization that hints at this improvement, yet this metric is still much higher than that of time-slicing or MIG, since SM provisioning does not isolate the rest of the memory subsystem as these workloads demand.

\begin{figure}[t!]
    \centering
    \begin{minipage}[t]{0.38\textwidth}
    \strut\vspace*{-\baselineskip}
        \centering
        \includegraphics[width=\linewidth]{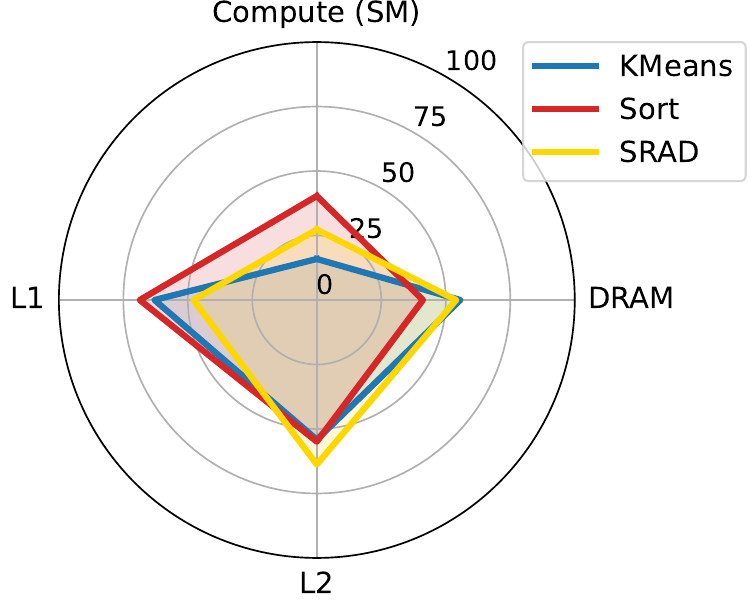}
        \captionsetup{
            format=plain,        
            indention=0pt,
            justification=raggedright,
            singlelinecheck=false
          }
        \caption{Throughput percentage (\%) profile  in NVIDIA A100\protect\footnotemark.}
        \label{fig:app_contention_prof}
    \end{minipage}
    \hfill
    \begin{minipage}[t]{0.58\textwidth}
        \centering
        \renewcommand{\arraystretch}{1.5}
        \captionsetup{
            format=plain,        
            indention=0pt,
            justification=raggedright,
            singlelinecheck=false
          }
        \captionof{table}{Duration (in seconds) and scaling efficiency according to the number of slices $s$ of the MIG instance ($E(s) = t(1) / (t(s) \cdot s)$).}
        \label{tab:MIG_times_contention}
        
\footnotesize
\setlength{\tabcolsep}{5.5pt}
\renewcommand{\arraystretch}{1.25}
\begin{tabular}{cccccccccc}\toprule

Slices & \multicolumn{2}{c}{KMeans} & & \multicolumn{2}{c}{Sort} & & \multicolumn{2}{c}{SRAD} \\
 \cmidrule{2-3} \cmidrule{5-6}
 \cmidrule{8-9}
$s$ & $t$ & $E$ & & $t$ & $E$ & & $t$ & $E$ \\
 \cmidrule{2-3} \cmidrule{5-6}
 \cmidrule{8-9}
1 & 16.27 & 1.00 & & 8.70 & 1.00 & & 4.65 & 1.00 \\
2 & 5.56 & 1.46 & & 4.23 & 1.03 & & 2.46 & 0.95 \\
3 & 4.58 & 1.18 & & 4.02 & 0.72 & & 1.48 & 1.05 \\
4 & 4.13 & 0.98 & & 3.95 & 0.55 & & 1.29 & 0.90 \\
7 & 3.68 & 0.63 & & 3.75 & 0.45 & & 0.96 & 0.69 \\
\hline
\end{tabular}
\label{tab:efficiency-slices}

        \strut\vspace*{-\baselineskip}
    \end{minipage}
\end{figure}
\footnotetext{NCU reports metrics at the kernel level, and each application consists of several kernels. To reduce the results to the application level, the metrics of its kernels have been weighted according to their duration.}

\begin{figure}[t]
    \centering
    \includegraphics[width=\linewidth]{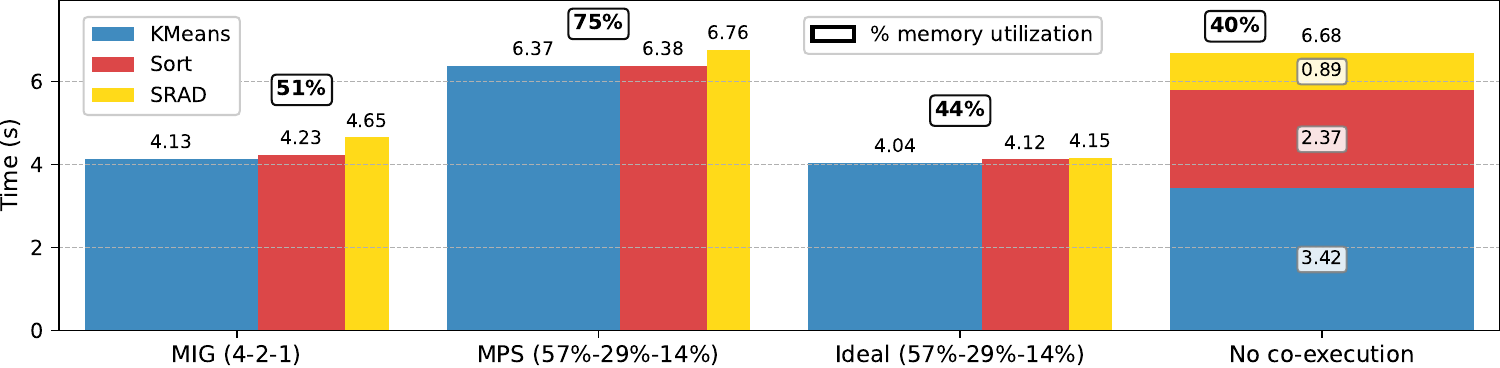}
    \caption{Duration and memory utilization on A100 for three benchmarks under MIG, MPS, non-interference ideal and no concurrency (bar width for the available resources).}
    \label{fig:MPS_contention_A100}
\end{figure}

To confirm that this memory contention is a fundamental issue and not limited to two-application tests on an NVIDIA A30, we evaluated additional co-execution patterns on an NVIDIA A100. The results were similar, demonstrating the significance of the previous results. For instance, we selected the M-type workloads \KMEANS, \SORT and \SRAD, which individually make little use of the throughput offered by the resources of an NVIDIA A100 (see \Cref{fig:app_contention_prof}), but mainly use memory (see \Cref{sec:metrics} for detail about NCU's throughput percentages). Based on their durations and efficiencies (\Cref{tab:efficiency-slices}), an effective co-execution would assign 4 slices to \KMEANS, 2 to \SORT, and 1 to \SRAD. In this sense, \Cref{fig:MPS_contention_A100} compares four scenarios based on this resource ratio: (1) co-execution with MIG using a 4-2-1 partition; (2) MPS with equivalent SM provisioning percentages ($\simeq$57\%, $\simeq$29\%, $\simeq$14\%); (3) an ideal case simulating no inter-application interference (execute each alone with the corresponding SM provisioning and assume that they would co-execute equally); and (4) sequential execution. The results of MPS are significantly worse than of MIG, with the percentage of time spent on memory operations increasing from 51\% to 75\%, which is consistent with our contention hypothesis. While applications under MIG finish considerably earlier than with MPS, they do not match the performance of the ideal interference-free case (MPS without real co-execution), which demonstrates that the problem is not MPS itself, but rather the simultaneous use of shared memory resources. In fact, MPS performs even worse than sequential execution in this experiment, despite the limited use of GPU throughput by applications. Furthermore, the MIG results are close to the ideal case, highlighting the suitability of this partitioning scheme for these tasks and indicating a low intrinsic overhead for the technology.

\subsection{Resource monopolization and fair sharing}

\begin{figure}[t]
    \centering
    \includegraphics[width=\linewidth]{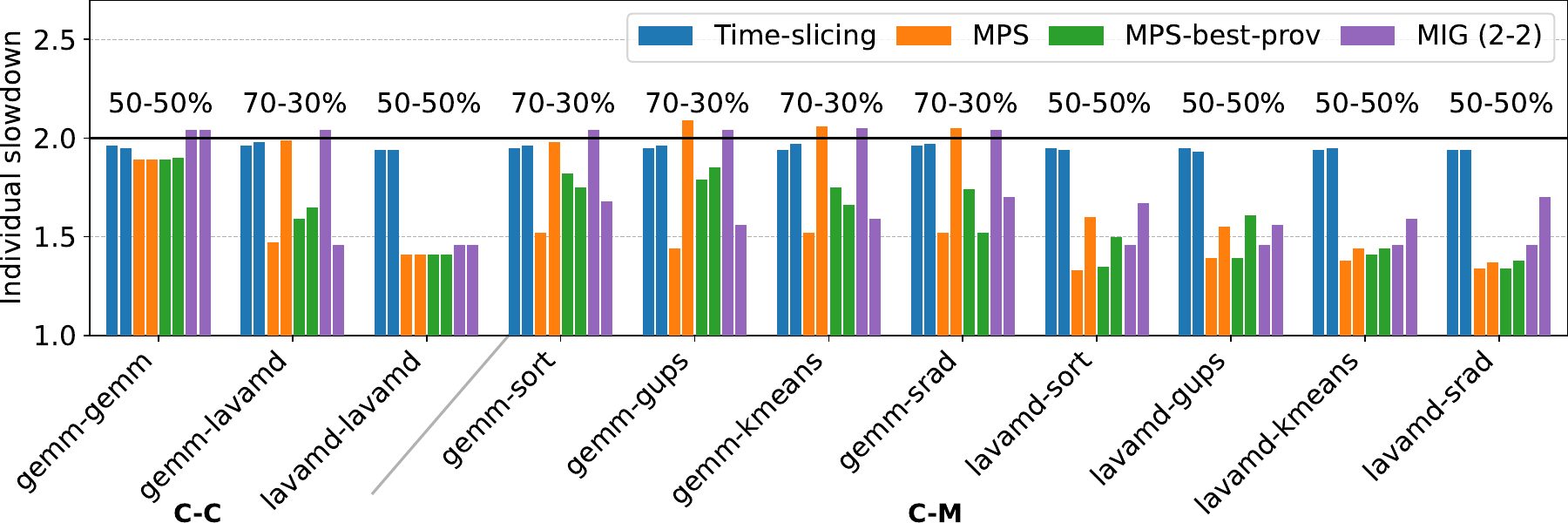}
    \caption{Co-execution slowdowns for pairs between C and M type workloads. Above the bars for each pair, the best percentages selected for MPS provisioning are indicated.}
    \label{fig:slowdown_CC_CM}
\end{figure}

\Cref{fig:slowdown_CC_CM} illustrates the slowdown for C-C and M-C workload combinations. In contrast to the MPS contention issues previously observed for memory-intensive (M-M) pairs, these new combinations show that MPS often achieves slightly better results than MIG and significantly better ones than time-slicing. This holds true even for \mbox{M-C} pairs, where one application has moderate memory usage but does not co-execute with another that heavily competes for those resources. These results reaffirm that the root cause of the problem is the shared memory contention within the MPS service. This also suggests that when such contention problems are absent, MPS can outperform MIG—since each application can access the full memory capacity. However, the performance differences observed in these cases are small and not very significant.

A striking observation from \Cref{fig:slowdown_CC_CM} is the poor performance of both MPS and MIG for one of the two applications in pairs involving the \GEMM benchmark. With MPS, the application co-executing with \GEMM is the one that experiences a significant slowdown, sometimes exceeding a factor of 2x (e.g., \GEMM with \GUPS or \SRAD). Conversely, with MIG, it is \GEMM itself that is penalized, while the co-running application is less affected. We attribute this to the fact that, as explained in \Cref{sec:meth:workloads}, \GEMM is a much more demanding application than the others, although mainly in terms of calculation rather than memory access. Consequently, under MPS, \GEMM monopolizes the SMs, preventing the co-running application from accessing them, even if that application could better utilize the underused memory system. With MIG, \GEMM is confined to an instance with half the GPU resources (one instance in a 2–2 partition), which causes  performance to degrade by more than double because its demand is higher. Meanwhile, the less demanding co-running application, isolated in its own 2-slice instance, is not similarly impacted (the amount of resources is sufficient for it).

To address this inefficiency, MPS provisioning, which allows limiting the percentage of SMs each process can use, proves effective. As shown in \Cref{fig:slowdown_CC_CM}, provisioning significantly improves the performance of the application running alongside \GEMM without excessively penalizing \GEMM itself (its slowdown remains well below 2$\times$). In our experiments, a 70-30\% SM split yielded the best results, which was determined through the parameter sweep mentioned in \Cref{sec:schemes}. However, the optimal configuration can vary, and should be guided by QoS requirements and a brief application profiling. The key takeaway is that the MPS daemon alone does not manage resource allocation efficiently---or at least, not fairly for the less demanding processes---but the problem can be effectively mitigated with the right choice of provisioning. Furthermore, MIG may struggle to provide sufficient resources due to its rigid partitioning limitations: an instance of 2 slices is insufficient for \GEMM, whereas a larger, 3-slice instance likely would be (as evidenced by its good performance with 70\% of SMs under provisioned MPS), but a 3-1 partitioning scheme is unfortunately not available in an NVIDIA A30.

\begin{figure}[t]
    \centering
    \includegraphics[width=\linewidth]{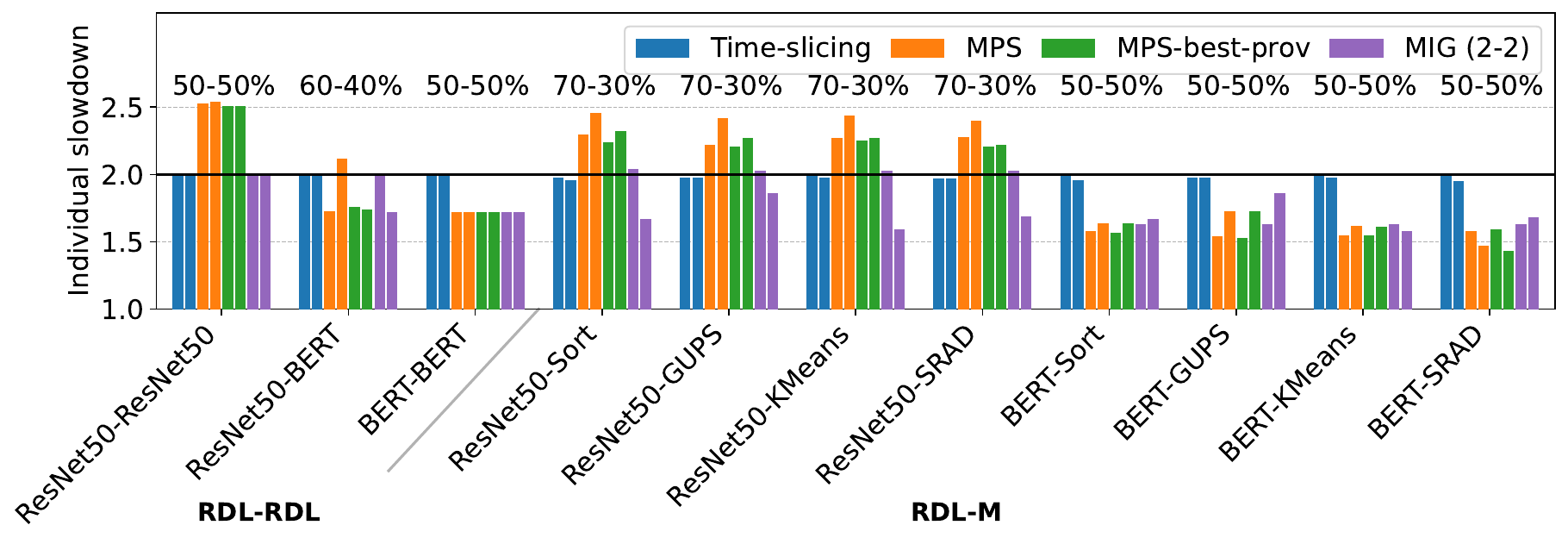}
    \caption{Co-execution slowdowns for pairs of RDL and M type workloads (the best MPS provisioning is indicated).}
    \label{fig:slowdowns_DRL}
\end{figure}

As mentioned, to enhance the comprehensiveness of the analysis, we included two deep learning applications that are representative of current trends and commonly used in GPU evaluations (\RESNETMEDIUM and \BERT). \Cref{fig:slowdowns_DRL} shows the slowdown for all possible pairs of these applications, and their combinations with M-type applications. For \RESNETMEDIUM, which was more intensive in both computation and memory (see \Cref{tab:app_metrics}), we observe the same phenomenon as when an M-type application is co-executed with another of the same type: presumably, MPS contention issues leading to performance losses of more than 2.5$\times$. This contrasts with \BERT, which, in the version used, is less demanding (see \Cref{tab:app_metrics}) and co-executes effectively with nearly all applications, similar to the C-type applications that resembled its usage profile. To a lesser extent, the performance of \RESNETMEDIUM does not improve with MIG (slowdown $\simeq 2x$) and affects the other application in MPS co-execution with \BERT, which is largely mitigated by optimal provisioning (a 60-40\% in this case). As was the case with \GEMM, \RESNETMEDIUM appears to monopolize resources with basic MPS and requires more than what a 2-slice MIG instance can offer. In essence, all previous observations are reinforced.

\subsection{Impact of partitioning granularity and flexibility}\label{sec:MIG_inflexibility}

As observed in Figures \ref{fig:slowdown_CC_CM} and \ref{fig:slowdowns_DRL}, the \GEMM and \RESNETMEDIUM applications suffer from significant performance degradation when co-executing on 2-slice instances (more than double the runtime). This issue would presumably not occur on a 3-slice instance (their co-execution performs well under MPS when provisioned with 70-30\% and 60-40\% of the SMs, respectively), but such a configuration is not feasible due to technology limitations. However, in these cases, it could be argued that co-locating applications is not as critical, as the problem arises from their already high resource utilization (see \Cref{tab:app_metrics}). This means MIG partitioning lacks the granularity and flexibility to accommodate such fine-grained adjustments.

However, the problem not only lies in the coarse-grained nature of the partitioning and the limited valid configurations, but also that every instance maintains the same proportion of different resource types (as explained in \Cref{sec:MIG_concepts}, instances group identical slices, so the ratio of SMs, memory capacity, or bandwidth, is always the same). This characteristic can be unsuitable for applications with asymmetric demand patterns (e.g., when the ratio between compute demand on the SMs and memory subsystem usage differs significantly from the fixed ratio of the instances). The least detrimental solution with MIG might be to use an instance that satisfies the maximum of these demands, leading to an implicit underutilization of other resources. However, sometimes even this allocation is not possible due to partitioning granularity and restrictions, whereas MPS memory sharing can provide the necessary flexibility to distribute this resource (provided there is no contention for it, which, as seen earlier, has an even worse impact).

\begin{table*}[t!]
\centering
\caption{Example of MIG inflexibility issues, using two applications and reporting throughput usage and durations in an NVIDIA A30.}
\label{tab:partition_problems}
\begin{threeparttable}

\footnotesize
\setlength{\tabcolsep}{4pt}
\begin{tabular}{@{}lccccccccccccc@{}}\toprule
  & \multicolumn{5}{c}{No co-execution}
  & & \multicolumn{5}{c}{MIG (2–2)}
  & & MPS (50\%–50\%)\tnote{(a)} \\
\cmidrule{2-6}\cmidrule{8-12}\cmidrule{14-14}
  Bench. & \multicolumn{4}{c}{Throughput (\%)} & Time (s) & & \multicolumn{4}{c}{Throughput (\%)} & Time (s) & & Time (s) \\
  \cmidrule{2-5} \cmidrule{8-11}
   & SMs & L1 & L2 & RAM & & & SMs & L1 & L2 & RAM & & & \\
   \midrule
    \textsc{HotSpot} & 38 & 42 & 2 & 0 & 42.75 & & 38 & 45 & 2 & 0 &  65.32 & & 60.32 \\
   \textsc{Raytracing} & 39 & 46 & 32 & 5 & 44.93 &  & 36 & 19 & \cellcolor[HTML]{FFD7D5} 46 & \cellcolor[HTML]{FFD7D5} 46 & 92.24 & &  61.45\\
   \midrule
   End time: & & & & & 87.68 & & & & & & 92.24 & & 61.45\\
\bottomrule
\end{tabular}
\begin{tablenotes}
\item[(a)] SM and memory performance was profiled using the NCU tool, which is not compatible with MPS, and there is no tool that provides similar data.
\end{tablenotes}

\end{threeparttable}
\end{table*}

To illustrate this, we used tests with a different profile: in particular, \textsc{HotSpot} (50 passes of the Rodinia suite benchmark \cite{Rodinia} on a 1024 square grid) and \textsc{Raytracing} (15 passes of the level 3 from Altis \cite{Altis}). \Cref{tab:partition_problems} reports their execution times and the percentage of peak throughput for the SMs and the memory hierarchy on an A30 GPU (see \Cref{sec:metrics}) under the following three schemes: sequential execution (the entire GPU for each benchmark), a 2-2 MIG partition, and the MPS equivalent (50-50\% of SMs). When run sequentially, both applications utilize nearly 40\% of the SMs' peak performance, suggesting that a 2-2 partition is reasonable from a compute perspective. However, their memory profiles are very different: \textsc{HotSpot} makes little use of memory (almost exclusively local to the L1 cache), whereas \textsc{Raytracing} intensively uses the shared L2 cache (32\% of throughput). This suggests that the applications should not conflict with each other, exhibiting a rather asymmetric device usage pattern between compute and memory. When co-executed with MIG, \textsc{Raytracing} appears to saturate its instance's memory resources, shifting a large portion of accesses to DRAM (46\% of throughput) but with low SM utilization (36\%). In contrast, with MPS, the memory is shared between the two applications with little interference, achieving better co-execution than MIG (61.45s vs. 92.24s) and completing both jobs faster than the sequential execution.

It is worth noting several nuances regarding this issue. First, we have illustrated this on an A30 GPU because its partitioning space is more limited: it only has 4 slices, 3 instance sizes (1, 2, and 4), and 5 possible partitions, making it unfeasible, for example, to assign 3 or 4 slices to the application with higher memory demand and 2 to the other. On bigger GPUs such as the A100, H100, or B200, which feature higher throughput and greater MIG granularity (7 slices, 5 possible instance sizes, and up to 19 partitions), this issue becomes less relevant because it is, a priori, easier for applications to adapt well to one of the available device partition proportions. Second, MIG includes Compute Instances (see \Cref{sec:MIG_concepts}), which allow memory to be shared among concurrent applications within the same instance without sharing the SMs. With these instances, it is possible to obtain results similar to those observed with MPS, although in general cases, this implies predicting and controlling interference to avoid contention problems, which is not always easy, as some scheduling proposals have noted \cite{Xiao2018, Zhang2024}. Finally, the ideal solution would be a MIG-like, but asymmetric, isolation: for example, being able to form an instance with 2 compute slices and 4 memory slices, all of them fully isolated. However, that seems infeasible or technically very difficult, at least in the short term; currently, MIG uses isolation at the GPC level, but it would likely require the ability to isolate resources at a lower, decoupled level.



\section{Efficiency and scalability analysis}\label{sec:energy_scalability}

\subsection{Energy consumption}

Another dimension of great relevance in the comparison between co-execution technologies is energy consumption. However, it is closely tied to the performance analysis just carried out: cases of memory contention or GPU monopolization with MPS will inevitably make energy consumption worse than other options or even worse than sequential execution without sharing the device (this was already the case in terms of performance). Similarly, co-executions of applications that proved very efficient (slowdown of about 1.5$\times$ for both) will very likely be very efficient in their energy consumption as well. Nevertheless, the energy analysis is not exactly the same as the performance analysis, since it is also influenced by the power consumed by the device (energy per unit time, i.e., watts), which in principle should increase with co-execution as device resources are used more intensively (in addition to control logic or other overheads introduced by the technologies).

\begin{figure}[t]
    \centering
    \includegraphics[width=\linewidth]{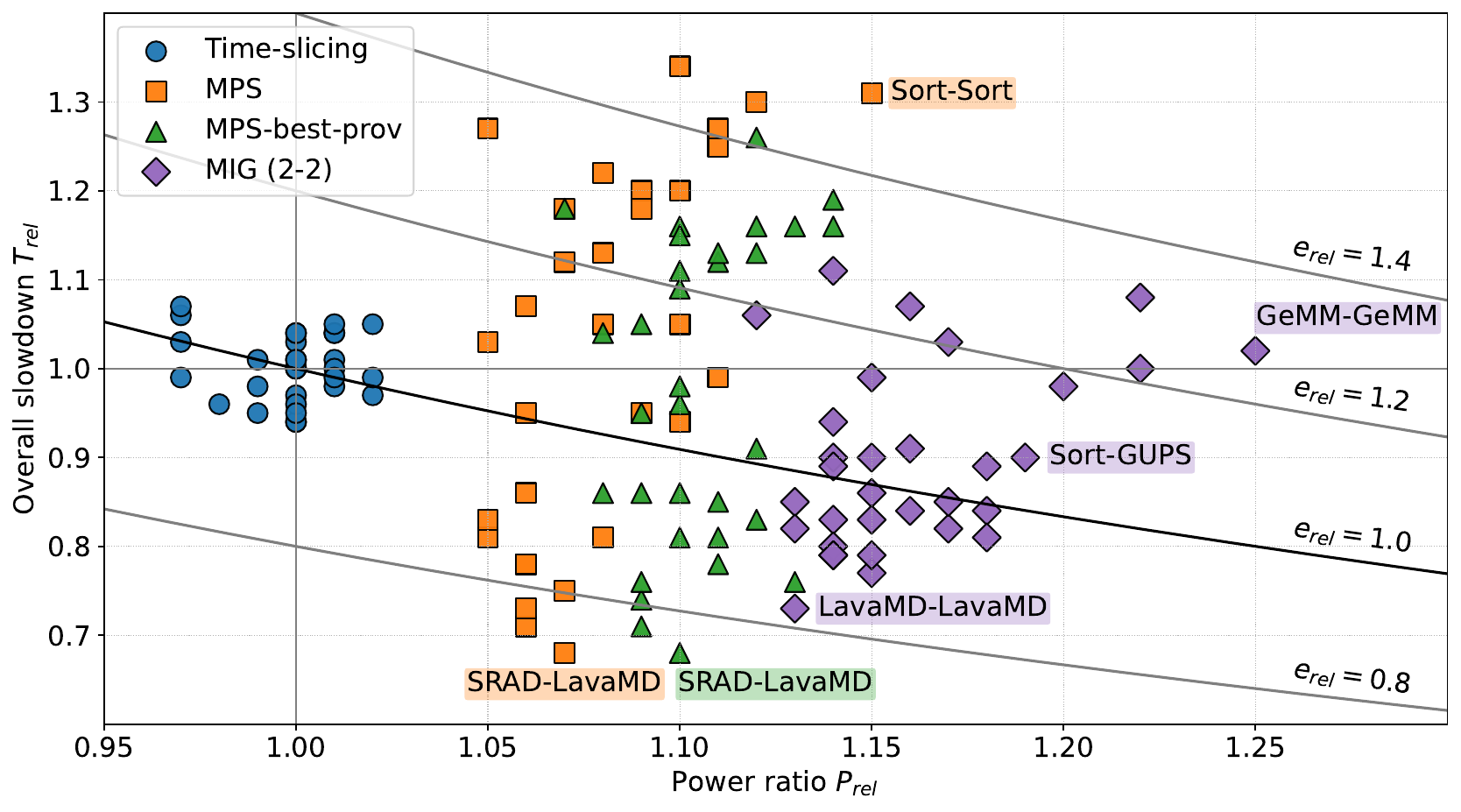}
    \caption{Scatter plot of application pairs according to their power increase ratio (x-axis $P_{\text{rel}} = P_{\text{co-exec}} \,/\, P_{\text{seq}}$) and their overall slowdown (y-axis $T_{\text{rel}} = T_{\text{co-exec}} \,/\, T_{\text{seq}}$). The isocurves for constant values of the energy ratio ($e_{\text{rel}} = x \cdot y$) are highlighted.}
    \label{fig:scatter_energy}
\end{figure}

To visualize these aspects together, \Cref{fig:scatter_energy} presents a scatter plot for all the application pairs executed on an NVIDIA A30 under each co-execution scheme (color legend). The x-axis represents the relative power draw, which is the ratio of the average power during co-execution to that of sequential execution: $P_{\text{rel}} = P_{\text{co-exec}} \,/\, P_{\text{seq}}$. The y-axis shows the overall slowdown (see \Cref{sec:metrics}), i.e., the ratio of the co-execution duration to the sequential execution time: $T_{\text{rel}} = T_{\text{co-exec}} \,/\, T_{\text{seq}}$.\footnote{Note that $T_\text{seq}$ and $P_\text{seq}$ are relative to the sequential execution of both applications, resulting in ratios close to 1. This differs from the slowdowns of \Cref{sec:performance}, where individual slowdowns were calculated to verify that the performance of one application did not degrade the other, so the values were around 2.} Notably, the product of these two axes corresponds directly to the relative energy consumption: $x \cdot y = (P_{\text{co-exec}} \cdot T_{\text{co-exec}}) \,/\, (P_{\text{seq}} \cdot T_{\text{seq}}) = E_{\text{co-exec}} \,/\, E_{\text{seq}}$. This relationship allows for plotting energy isocurves for key values, where a constant energy ratio is defined by $e_{\text{rel}} = x \cdot y$. The plot includes a primary isocurve for a ratio of 1.0, which serves as a baseline: points below this line represent a net reduction in energy consumption, while points above it indicate an increase. In addition, consumption isocurves at 0.8 (20\% reduction), 1.2 (20\% increase), and 1.4 (40\% increase) are also highlighted. Finally, in the points we will refer to, the name of the co-executed pair is highlighted.

As observed in \Cref{fig:scatter_energy}, the time-slicing scheme maintains a power draw very close to the sequential baseline (points are clustered around $P_{\text{rel}}\!=\!1$). This is expected, as this method does not allow for the simultaneous use of functional units (only alternation). Under time-slicing, the total energy consumption improves for some application pairs and degrades for others (falling below or above the isocurve, respectively), primarily depending on the slight performance variations discussed previously (the y-axis value). However, these changes are modest, with no application pair approaching the 0.8 or 1.2 isocurves.

In contrast, spatial sharing methods increase power draw more significantly (which is expected due to their higher use of resources), but in many cases this is cost-effective, as performance improves enough to reduce overall energy consumption (by up to 20\% in some cases). In \Cref{fig:scatter_energy}, MPS shows power increases of around 7\% for pairs without contention or resource monopolization issues, such as \SRAD-\LAVAMD ($P_{\text{rel}}\!=\!1.07$, $T_{\text{rel}}\!=\!0.68$). This increase reaches up to 15\% for pairs that do exhibit these problems, like \SORT-\SORT (an M-M combination with memory contention and $P_{\text{rel}}\!=\!1.15$, $T_{\text{rel}}\!=\!1.31$). These wide performance variations, coupled with the differing power draws, lead to highly variable energy outcomes. This results in cases with over 20\% energy improvement, such as \SRAD-\LAVAMD (below the 0.8 isocurve), and others like \SORT-\SORT with a degradation of over 40\% (above the 1.4 isocurve). The conclusion is that co-execution with MPS can yield extreme outcomes, both positive and negative. It is therefore critical to avoid contention scenarios, a conclusion that is even more pronounced for energy consumption than it was for performance.

Using MPS with SM provisioning results in a consistently higher power draw versus basic MPS, ranging from 10\% to 15\% extra. This is potentially attributable to the additional logic required for SM-level isolation. However, there appears to be no significant difference in the overall power-performance trade-off, as the data points for both basic MPS and MPS-best-prov show a similar distribution relative to the isocurves (see particularly \SRAD-\LAVAMD, which is highlighted in both). This suggests that the higher power draw of provisioning is generally offset by corresponding performance gains.

Finally, MIG exhibits the highest power levels in \Cref{fig:scatter_energy}: typically around a 15\% increase, reaching 25\% when handling applications more demanding than the assigned instance such as \GEMM-\GEMM. This is natural, because full resource isolation must necessarily incur an energy overhead, especially when the instance size is insufficient for the application's demand as in the \GEMM case and causes issues like increased stress on high memory levels (as in the example from \Cref{tab:partition_problems}). Although MIG delivered the best average performance (with y-values more frequently below 1.0), its peak performance does not match the best cases of MPS, and conversely, it also avoids severe performance degradation. Consequently, the energy trade-off is many times unfavorable, as approximately half of the MIG test cases fall above the 1.0 isocurve, even though most are below $T_\text{rel} = 1$ (they improve the original performance). For example, \SORT-\GUPS, despite reducing execution time by 10\% ($T_{\text{rel}}\!=\!0.9$), is not energy-efficient because its power draw increases by 19\% ($P_{\text{rel}}\!=\!1.19$). Furthermore, even in cases where MIG does yield energy savings, these savings do not reach the levels seen with MPS (no MIG pair falls below the 0.8 isocurve, although LavaMD-LavaMD is close). The conclusion is that MIG hardware isolation carries an important inherent energy cost, and to be energy efficient, the performance gains must be substantial enough to offset this overhead (which occurs in approximately half of our test cases).

\subsection{Performance concurrency scaling}

The detailed analysis of each combination has so far allowed us to characterize key aspects of the technologies, but to do so, the degree of concurrency has been kept fixed at two applications (pairs of workloads, in order to analyze each case). In this regard, we extend the study in this dimension by also evaluating performance scaling to four concurrent applications (the MIG instance not evaluated so far on an A30 GPU is the 1-slice instance, which allows up to four applications to run on a 1-1-1-1 partition). Since the possible combinations grow exponentially with the number of applications (the reason why an exhaustive analysis was only done with 2 applications), we limit the co-execution to copies of the same application (in particular, these combinations should reflect the MPS contention issues that occurred between M-type workloads). Naturally, the scaling will be conditioned by the applications' resource utilization; while co-executing pairs of these workloads was reasonable based on the profiles of \Cref{tab:app_metrics}, running four of them simultaneously would be unrealistic. Consequently, we divide this study into two scenarios:
\begin{itemize}
    \item \uline{Medium workloads}: Maintaining the same configuration as in the previous tests, stated in \Cref{tab:workloads_specification}.
    \item \uline{Small workloads}: Compared to the specification in \Cref{tab:workloads_specification}, the workload level was reduced by one for Altis benchmarks, \KMEANS was run on only 16,384 points with 16 dimensions and 16 centroids, and the RDL models were replaced with smaller equivalent versions: \RESNETSMALL \cite{ResNet} and \DISTILBERT \cite{DistilBERT}. \Cref{tab:small_apps} shows the metrics of new durations, utilization, and energy for these tests, which are now very undemanding (except for \GEMM, whose utilization is moderate).\footnote{In this case, it is not as important for the durations to be similar because we will not be co-executing different tests, but rather multiple copies of the same application.}
\end{itemize}

\begin{table*}[t!]
\centering
\caption{Small workloads metrics when running solo on an NVIDIA A30.}
\label{tab:small_apps}

\footnotesize
\renewcommand{\arraystretch}{1.1}
\setlength{\tabcolsep}{3pt}
\begin{tabular}{cccccccccc}\toprule
 & & & \multicolumn{2}{c}{SMs utilization (\%)} & & \multicolumn{2}{c}{Mem. utilization (\%)}
  & & \\
\cmidrule{4-5} \cmidrule{7-8}
 Type & Test name & Time (s) & Avg. & Max. & & Avg. & Max. & Energy (J) & Power (W) \\ \midrule
  M & \SORT & 2.84 & 12.20 & 20 & & 9.42 & 22 & 140.36 & 49.42 \\
  M & \GUPS & 1.11 & 7.18 & 15 & & 12.11 & 58 & 52.28 & 47.10 \\
  M & \KMEANS & 0.53 & 6.07 & 12 & & 8.22 & 20 & 24.00 & 45.28 \\
  M & \SRAD & 2.12 & 8.87 & 20 & & 6.44 & 32 & 92.12 & 43.45 \\ \midrule
  C & \GEMM & 2.54 & 35.24 & 100 & & 1.20 & 28 & 187.88 & 73.97 \\ 
  C & \LAVAMD & 0.24 & 5.09 & 49 & & 0.51 & 5 & 10.20 & 42.50 \\ \midrule
  RDL & \RESNETSMALL & 2.16 & 10.11 & 42 & & 2.21 & 22 &  124.21 & 57.50 \\
  RDL & \DISTILBERT & 1.82 & 12.52 & 30 & & 3.39 & 45 & 83.56 & 45.91 \\
\bottomrule
\end{tabular}
\end{table*}

\begin{figure}[t!]
    \centering
    \begin{subfigure}[b]{\textwidth}
        \centering
        \includegraphics[width=\textwidth]{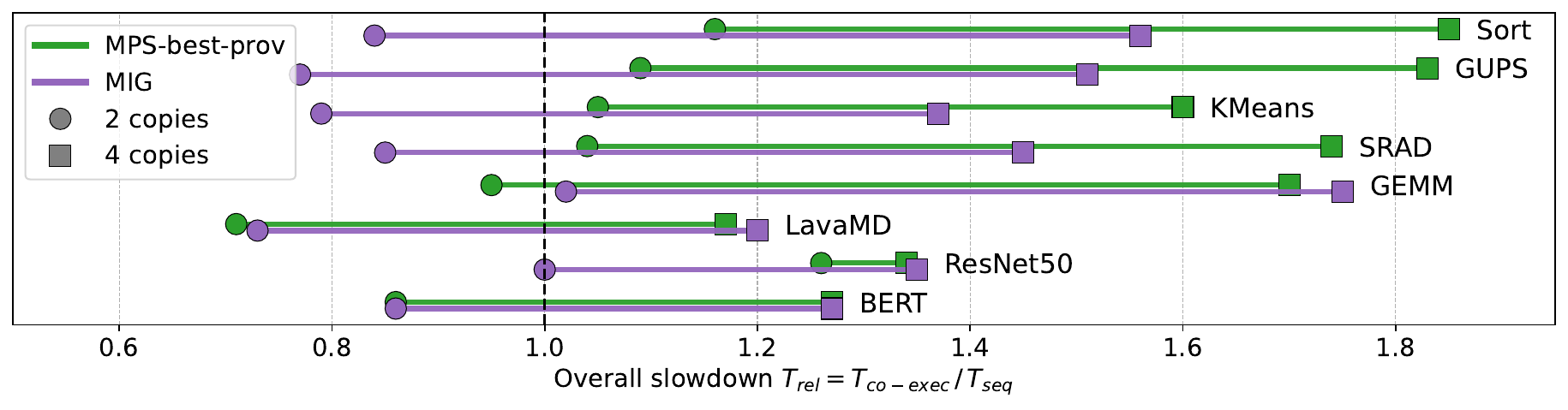}
        \caption{Medium workloads.}
        \label{fig:perf_scaling_medium}
    \end{subfigure}
    \begin{subfigure}[b]{\textwidth}
        \centering
        \includegraphics[width=\textwidth]{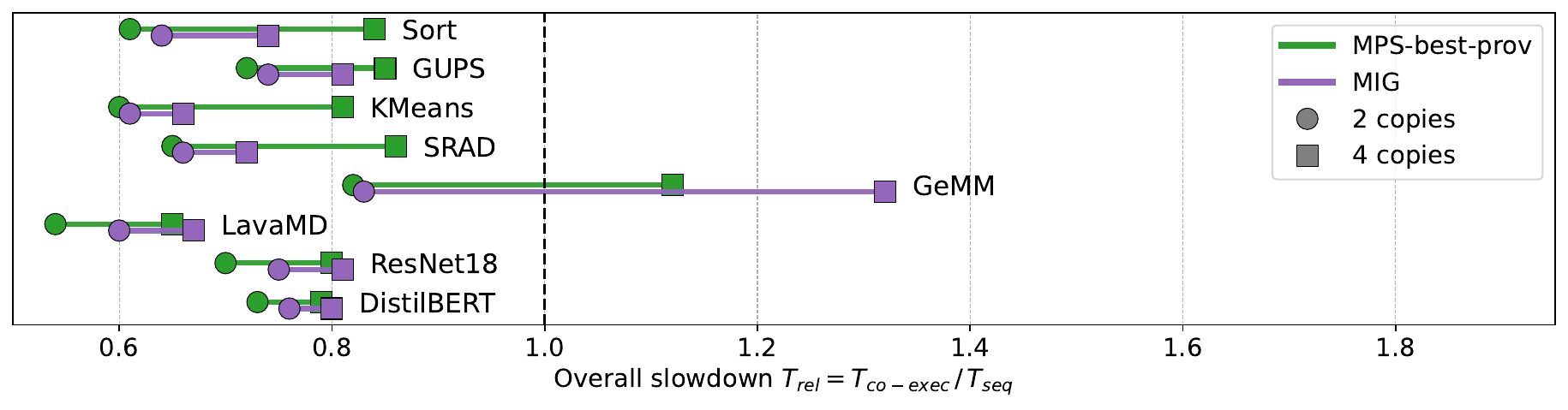}
        \caption{Small workloads.}
        \label{fig:perf_scaling_small}
    \end{subfigure}

    \caption{Dumbbell plots comparing the overall slowdown when co-executing 2 versus 4 copies of each application under different co-execution schemes.}
    \label{fig:perf_scaling}
\end{figure}

To streamline the analysis, we exclude time-slicing from the results, as it offers nothing new: as previously noted, it behaves like sequential execution in all cases. We also stick with the MPS-best-prov scheme, which in all cases outperforms MPS, and it is very easy to determine the optimal provisioning beforehand as copies of the same application are co-executed, so SMs must be distributed uniformly: 50-50\% for 2 applications and 25-25-25-25\% for 4. \Cref{fig:perf_scaling} shows the changes in overall slowdown when moving from 2 to 4 concurrent copies of each application, comparing both schemes and workload sizes (Medium and Small).

For the Medium workloads (\Cref{fig:perf_scaling_medium}), the poor scaling is striking: performance degrades significantly compared to sequential execution, showing a wide and unfavorable shift in overall slowdown. This is logical, as the utilization profiles of these tasks suggested that limiting their resources to a quarter of what is available on this GPU would be insufficient. For M-type applications (\SORT, \GUPS, \KMEANS, and \SRAD), this effect is much more exaggerated with MPS, likely due to the memory contention issues being greatly exacerbated by having more applications sharing resources in parallel. With MIG, this degradation is more moderate for those applications due to its resource isolation. In any case, it can be concluded that the degree of application co-execution must be closely linked to their level of resource underutilization, otherwise the performance losses can be huge (over 80\% slower than sequential execution in some cases).

In contrast, the Small workloads also show a worse overall slowdown when moving from 2 to 4 concurrent copies (\Cref{fig:perf_scaling_small}), but in a much more moderate fashion. They start from more notable improvements over sequential execution (overall slowdowns around 0.6 in many cases) and are still preferable to no co-execution even with 4 applications (a small increase in slowdown that remains well below 1.0). The exception is \GEMM, which remains relatively demanding even in its Small version, making it inefficient to co-execute 4 copies in performance terms (much like many Medium workloads). It is also apparent for the M-type applications (the first four) that scaling to 4 copies is considerably better with MIG than with MPS-aprov (this is not the case for the other workloads). Again, this appears to be a consequence of MIG's total isolation versus MPS's shared memory system. In conclusion, if resource underutilization is higher, opting for a finer-grained MIG partition to co-execute more applications seems to be a good option, especially if their memory system usage is not entirely residual.

\subsection{Energy efficiency scaling}

\begin{figure}[t]
    \centering    \includegraphics[width=\linewidth]{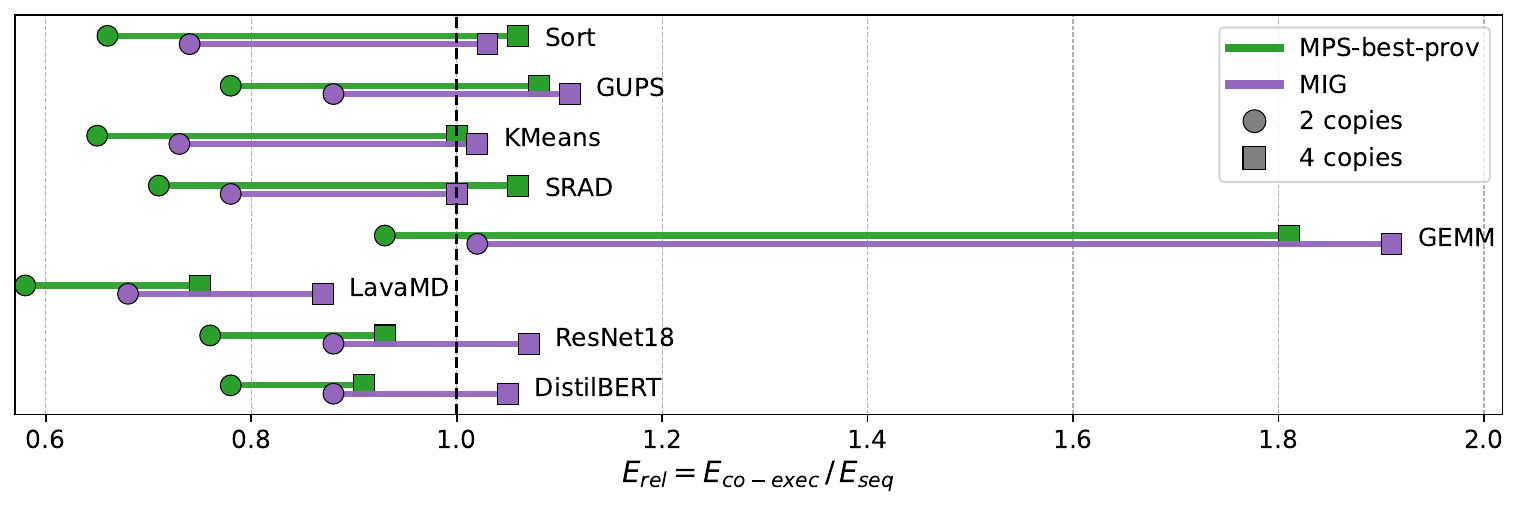}
    \caption{Energy consumption ratio from 2 to 4 concurrent copies of Small workloads.}
    \label{fig:energy_scaling}
\end{figure}

\begin{figure}[t]
    \centering    \includegraphics[width=\linewidth]{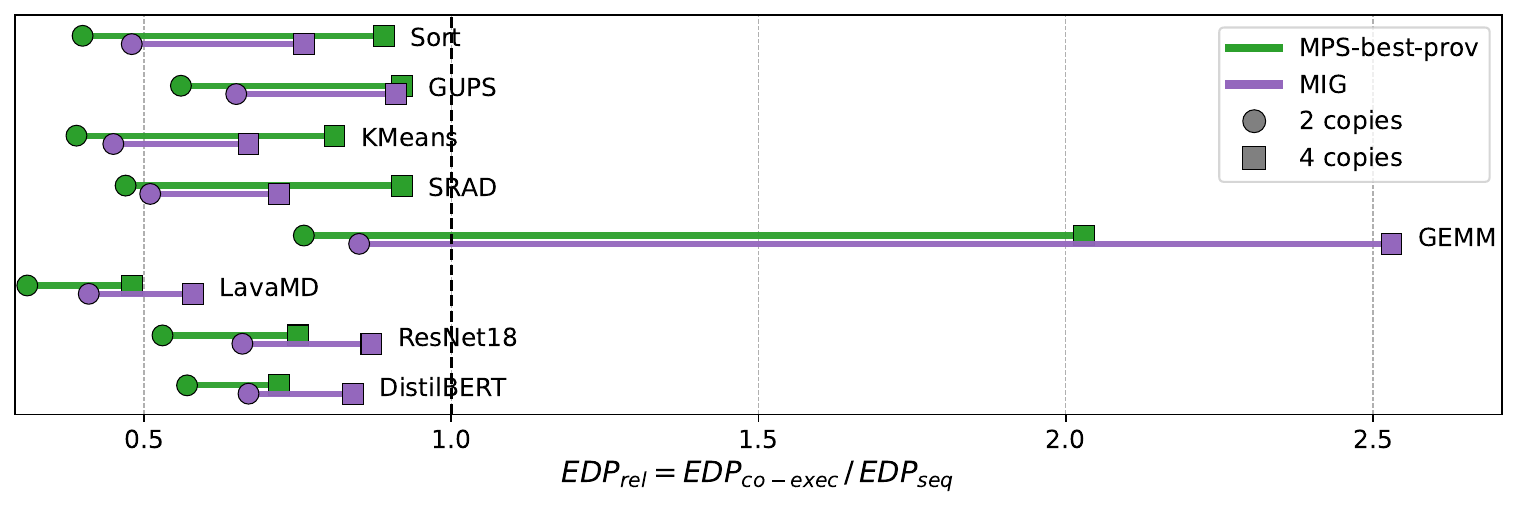}
    \caption{Energy-Delay Product ratio from 2 to 4 concurrent copies of Small workloads.}
    \label{fig:edp_scaling}
\end{figure}

To analyze energy consumption scaling, we focused on the Small applications. As shown in \Cref{fig:perf_scaling_medium}, running four concurrent instances of the Medium workloads significantly worsens performance and, consequently, energy consumption (given this clear inefficiency, a deeper analysis of this scenario was not pursued). For the Small applications, we analyzed the variation in the relative energy increase ratios $E_{\text{rel}}$ when moving from two to four concurrent applications (\Cref{fig:energy_scaling}). To better evaluate the trade-off with performance, we also examined the relative increase ratios of the Energy-Delay Product (EDP) in \Cref{fig:edp_scaling}, defined as the energy consumed multiplied by the execution time: $EDP_{\text{rel}} = E_{\text{rel}} \cdot T_{\text{rel}} = (E_{\text{co-exec}} \cdot T_{\text{co-exec}})\,/\,(E_{\text{seq}} \cdot T_{\text{seq}})$.

The energy consumption results in \Cref{fig:energy_scaling} accentuate the inefficiencies of both schemes when scaling to four applications. While running two concurrent copies generally consumes less energy than sequential execution, scaling to four copies often proves detrimental (all of them except \LAVAMD and \GEMM). In fact, this is purely due to energy overhead because these applications were faster than the reference with four copies ($T_{\text{rel}} < 1$ in \Cref{fig:perf_scaling_small}). In the case of the M-type applications, the effects of memory contention that occur with MPS but not MIG are accentuated to the point that the consumption of four copies of \SORT, \GUPS, \KMEANS, or \SRAD is greater with MPS than with MIG, but it was the opposite with two copies. In contrast, for applications without contention issues like \RESNETSMALL and \DISTILBERT, MIG's overhead leads to significantly higher energy use, even surpassing the sequential baseline ($E_{\text{rel}} > 1$), while MPS remains more efficient ($E_{\text{rel}} < 1$). One clear conclusion is that the effect of both phenomena intensifies significantly as the degree of concurrency increases. Along the same lines, for \GEMM, a significantly more demanding application than the others, co-executing four copies is extremely detrimental, increasing energy consumption by approximately 80\% compared to its sequential baseline. This underscores the importance of profiling application demands to determine an appropriate degree of concurrency. Conversely, an application like \LAVAMD, which saw substantial performance gains (\Cref{fig:perf_scaling_small}), continued to improve its energy efficiency at four copies with both schemes.

Regarding the Energy-Delay Product (EDP), the results in \Cref{fig:edp_scaling} show that performance improvement successfully compensate for the increased energy consumption with four concurrent copies. As shown in \Cref{tab:small_apps}, the reduction in execution time was approximately 20\% compared to the sequential baseline, while the corresponding energy increase was only around 10\% (see in \Cref{fig:energy_scaling} \SORT, \GUPS, \KMEANS and \SRAD, especially with MPS, and \RESNETSMALL and \DISTILBERT with MIG). Consequently, the overall EDP, which gives equal weight to both factors, is favorable ($EDP_{\text{rel}} < 1$),  indicating that the gains in performance outweigh the energy penalty (for all application except \GEMM). The application-by-application analysis reveals a combination of the phenomena already discussed, and it is striking how \GEMM's EDP is extremely poor with 4 copies (not so with 2, with an enormous difference between them) because its demand is excessive for that degree of concurrency, which impairs performance but also increases consumption (unlike other tests where compensatory trade-offs were observed).

\section{Related work}\label{sec:related_work}

While the growing underutilization of GPU resources has led to significant interest in co-execution technologies like MPS and MIG \cite{Adufu2024, Durvasula2024}, prior work has predominantly centered on developing job schedulers to optimize a target metric \cite{Li2022_MISO, VillarrubiaFAR2025, VillarrubiaDRL2026, Zhang2024}. In these studies, the evaluation of the technologies themselves is often secondary and limited, serving merely as a preliminary motivation for their proposed techniques rather than the focus of the research. Consequently, these analyses lack the depth required for a thorough understanding. For instance, works like MISO \cite{Li2022_MISO} and MIGER \cite{Zhang2024} illustrate the memory contention issues in MPS versus the isolation in MIG, but their evidence is typically limited to a few DL models in a single provisioning configuration. They do not provide broader insights into the frequency of these problems or their correlation with different workload types. Similarly, while MIGER \cite{Zhang2024} explores technology scaling with concurrent applications, it does so with a few examples aimed only at motivating its optimization algorithms. In contrast, our study provides a comprehensive characterization that can serve as a source of lessons for future research.

Although considerably scarcer, some studies do focus purely on evaluating these technologies. However, they often have different scopes or limitations, such as focusing on a single technology rather than a direct comparison. For example, Wende et al. \cite{Wende2014} evaluated the underlying hardware support for MPS (Hyper-Q) at the level of OpenMP parallel regions and MPI processes, concluding that the process-level parallelism is much more effective (the approach used by MPS). Meanwhile, the work of Weaver et al. \cite{Weaver2024} is more analogous to ours, evaluating MPS with ten workload combinations characterized by their performance and energy efficiency. Nonetheless, their study does not utilize MPS resource provisioning nor does it compare MPS against MIG, which are two fundamental axes of our work. Other studies like MIGPerf \cite{Zhang2023} focus primarily on MIG, and concentrate on DL inference using batch size as a primary scaling dimension. Finally, Robroek et al. \cite{Robroek2024} provide a direct comparison between MPS and MIG with a larger scaling test than in our study (up to 7 concurrent instances on an A100), but they do not use the MPS provisioning option and evaluate three pairs of DL workloads that do not incur memory contention.

Our study focuses on MPS and MIG as they are the standard GPU sharing technologies provided by the NVIDIA driver (although MIG is only supported by some GPU models). Unlike experimental frameworks, MPS and MIG are integrated into the hardware and driver ecosystem, making their performance a critical baseline for real-world deployments. Understanding their overhead and limitations is therefore more relevant for practitioners than analyzing research prototypes. However, other lower-level works have presented custom concurrency tools that explicitly manage CUDA streams and application switching, usually oriented to more specific goals and workloads. For example, Tally \cite{Zhao2025} is a virtualization layer that provides performance isolation, reporting a 7.2\% average overhead on inference latency, compared to 195.5\% for MPS and 188.9\% for the prior state-of-the-art \cite{Wu2023TSC}.
\section{Conclusions}\label{sec:conclusions}

This paper has presented a comprehensive empirical evaluation of spatial co-execution on modern GPUs, focusing on NVIDIA's MPS and MIG technologies. We have compared these against a standard time-slicing baseline, analyzing their performance, energy consumption, and scalability across a diverse set of workloads with varying computational and memory demands. Our findings highlight the significant potential of spatial sharing to improve GPU utilization and efficiency, while also revealing the critical trade-offs between the flexibility of MPS and the strong isolation of MIG. The key insights obtained are summarized below.
\subsection{Key insights}
A primary finding is that, when correctly matched with workloads, both spatial sharing technologies can yield substantial improvements over the standard time-slicing baseline. In some of our test cases, this resulted in overall performance improvements of up to 30\% ($T_{\text{rel}} \!\simeq \!0.7$ in \Cref{fig:scatter_energy}) and reductions in energy consumption approaching 20\%. However, the path to achieve these gains critically depends on understanding the trade-offs of each technology.

\begin{itemize}[leftmargin=0pt]
    \item[\ding{226}] \textit{Co-execution through MIG hardware resource isolation}\,\\[0.4\baselineskip]
    \indent MIG provides robust performance isolation by partitioning hardware resources, which effectively prevents interference between co-running applications, albeit with a minor overhead. Consequently, most application pairs achieved a notable performance improvement compared to sequential execution, with many showing a 20\% reduction in overall slowdown and some reaching up to 30\% (see \Cref{fig:scatter_energy}). This benefit was particularly stark for combinations of two memory-intensive (M-type) workloads, which completed their execution almost twice as fast as with MPS
    (see \Cref{fig:slowdown_MM}).
    
    As a drawback, \Cref{sec:MIG_inflexibility} highlighted that rigid, discrete partitioning of GPU resources into predefined instance sizes can lead to underutilization if an application's resource requirements do not align well with the available instance configurations (we saw in \Cref{tab:partition_problems} much slower co-execution with higher L2 and DRAM memory usage). The issue is especially pronounced on GPUs with fewer slicing options, such as the A30, where the lack of granularity can yield suboptimal performance. Furthermore, while MIG consistently improves energy efficiency over sequential execution, it introduces slightly more overhead than MPS and does not reach the same peak efficiency as the most favorable cases of MPS ($E_{\text{rel}} < 0.8$ in \Cref{fig:scatter_energy}). This efficiency gap became more noticeable when scaling to four concurrent applications (Figures \ref{fig:energy_scaling} and \ref{fig:edp_scaling}).\\[-0.5\baselineskip]

    \item[\ding{226}] \textit{MPS flexibility and memory sharing issues}\,\\[0.4\baselineskip]
    \indent MPS offers a more flexible approach to resource sharing, allowing for fine-grained allocation of computational resources. This flexibility, however, comes at the cost of potential contention for shared resources, particularly the memory subsystem. The ability to provision SMs provides a mechanism to mitigate some SM-level sharing issues and, crucially, to prevent resource monopolization, as was observed with the dominant \GEMM application in \Cref{fig:slowdown_CC_CM}. Nevertheless, this provisioning does not solve the underlying memory contention problems inherent in memory-intensive (M-type) applications. As a result, the performance and energy consumption outcomes with MPS are highly polarized (see \Cref{fig:scatter_energy}). It achieves better results than MIG in favorable scenarios but performs significantly worse when memory conflicts arise. This polarization was further accentuated when scaling to four concurrent applications (Figures \ref{tab:small_apps} and \ref{fig:energy_scaling}). These findings underscore the critical importance of careful workload characterization to select the appropriate technology.\\[-0.5\baselineskip]
    
    \item[\ding{226}] \textit{Potential future enhancements}\,\\[0.4\baselineskip]
    \indent The evolution of MIG would benefit from more granular and flexible partitioning schemes to reduce resource underutilization. Specifically, the ability to create instances with asymmetric resource (e.g., more memory bandwidth relative to compute resources) would better match diverse application profiles. While technically challenging in the short term, this would help bridge the gap between MIG's strong isolation and MPS's flexibility.

    On the other hand, to mitigate its primary weakness, MPS could incorporate lightweight memory management mechanisms. While full isolation remains the domain of MIG, adding features that throttle memory bandwidth at the process level (analogous to SM provisioning) could prevent worst-case contention scenarios and make MPS a more robust solution across a wider range of applications.
\end{itemize}

\subsection{Future work}

Our findings pave the road for a dynamic, workload-aware scheduler that automatically selects the optimal co-execution strategy. While prior work has already explored this approach---particularly for MIG---our analysis provides a solid empirical foundation for developing more robust, and potentially hybrid, schedulers that combine MPS and MIG according to task characteristics.

\section*{Declarations}

\subsection*{Ethical Approval}

Not applicable.

\subsection*{Competing interests}

There are no competing interests.

\subsection*{Authors' contributions}

J.V. conducted the design, implementation and evaluation of the experiments, and wrote the main manuscript text.
L.C., F.I and K.O collaborated in the definition and supervision of research tasks, the design and critical analysis of the experiments, and the review and writing of the manuscript.

\subsection*{Acknowledgements}

This work is funded by Grants {PID2021-126576NB-I00} and {PID2024-158311NB-I00} funded by {MCIN/AEI/10.13039/5011\allowbreak00011033} and by {{\em ``ERDF A way of making Europe''}}. 
We thank the HPC\&A group at Universitat Jaume I de Castell\'on for granting us access to A100 and H100 GPUs for profiling and evaluation purposes.

\subsection*{Availability of data and materials}

Data sharing not applicable to this article as no datasets were generated or analysed during the current study.

%
%






\bibliography{biblio}

\end{document}